\documentclass[11pt]{article}
\usepackage{jcappub}

\usepackage{bm}
\usepackage{color}
\usepackage{graphicx}
\newcommand{\be}{\begin{equation}}
\newcommand{\ee}{\end{equation}}
\newcommand{\een}{\end{subequations}}
\newcommand{\ben}{\begin{subequations}}
\newcommand{\beq}{\begin{eqalignno}}
\newcommand{\eeq}{\end{eqalignno}}
\newcommand{\aq}{a_q}
\newcommand{\bq}{b_q}
\newcommand{\aQ}{a_Q}
\newcommand{\bQ}{b_Q}
\newcommand{\mo}{micrOMEGAs}
\newcommand{\Mo}{MicrOMEGAs}
\newcommand{\msq}{\mbox{$m_{\tilde{q}}$}}
\newcommand{\mc}{\mbox{$m_{\chi}$}}
\newcommand{\mqsq}{\mbox{$m^2_q$}}
\newcommand{\mcsq}{\mbox{$m^2_{\chi}$}}
\newcommand{\msqsq}{\mbox{$m^2_{\tilde{q}}$}}
\newcommand{\rt}{\mbox{$\sqrt{|\Delta|}$}}
\newcommand{\lsim}{\mathrel{\mathop{\kern 0pt \rlap
      {\raise.2ex\hbox{$<$}}}\lower.9ex\hbox{\kern-.190em $ \sim$}}}
\newcommand{\gsim}{\mathrel{\mathop{\kern 0pt
      \rlap{\raise.2ex\hbox{$>$}}}\lower.9ex\hbox{\kern-.190em $\sim$}}}
\newcommand{\coeffs}{$f^{(Q)}_G$, $f_q$, $g^{(i,Q)}_G$, $g^{(i)}_{q}$}
\newcommand{\minabs}{\mathop{\rm minabs}}

\title{On the sbottom resonance in dark matter scattering}

\author[a]{Paolo Gondolo,}
\author[b]{Stefano Scopel}
\emailAdd{paolo.gondolo@utah.edu}
\emailAdd{scopel@sogang.ac.kr}
\affiliation[a]{Department of Physics and Astronomy, University of Utah, Salt Lake City, Utah 84112-0830, USA}
\affiliation[b]{Department of Physics, Sogang University, Seoul, South Korea}

\abstract{
A resonance in the neutralino--nucleus elastic scattering cross
section is usually purported when the neutralino-sbottom mass
difference $m_{\tilde{b}} - m_{\chi} $ is equal to the bottom quark
mass $m_b \sim 4$ GeV. Such a scenario has been discussed as a viable
model for light ($\sim 10$ GeV) neutralino dark matter as explanation
of possible DAMA and CoGeNT direct detection signals. Here we give
physical and analytical arguments showing that the sbottom resonance
may actually not be there. In particular, we show analytically that
the one--loop gluon--neutralino scattering amplitude has no pole at
$m_{\tilde{b}}=m_{\chi}+m_b$, while by analytic continuation to the
regime $m_{\tilde{b}}<m_{\chi}$, it develops a pole at
$m_{\tilde{b}}=m_{\chi}-m_b$. In the limit of vanishing gluon momenta,
this pole corresponds to the only cut of
the neutralino self-energy diagram with a quark and a squark running
in the loop, when the decay process $\chi\rightarrow\tilde{Q}+Q$
becomes kinematically allowed. The pole can be interpreted as  the formation of a $\tilde{b}\overline{b}qqq$ or $\tilde{b}^* b qqq$ resonant state (where $qqq$ are the nucleon valence quarks), which is however kinematically not accessible if the neutralino is the LSP.  Our analysis shows  that the common practice of estimating the neutralino-nucleon cross section by introducing an ad-hoc pole at $m_{\tilde{b}}=m_{\chi}+m_b$ into the effective four--fermion interaction (also including higher--twist  effects) should be discouraged, since it corresponds to adding a  spurious pole to the scattering process at the center-of-mass energy  $\sqrt{s}\simeq m_{\chi}\simeq m_{\tilde{b}}-m_b$.  Our considerations can be extended from the specific case of supersymmetry to other similar cases in which the dark matter particle  scatters off nucleons through the exchange of a $b$--flavored state almost degenerate in mass with the dark matter particle, such as for instance in theories with extra dimensions and in other mass--degenerate dark matter scenarios recently discussed in the literature.
}

\begin{document}

\maketitle

\section{Introduction}
\label{sec:I}

With the inception of the LHC (Large Hadron Collider) operations in
2010, the reckoning time has finally come for supersymmetry (SUSY) and
other theories at the electroweak scale devised to solve the
naturalness problem of the Standard Model. Before the LHC shut-down at the end
of 2012 to prepare it for the upgrade to the final designed
center--of--mass energy of $\sqrt{s} = 14$ TeV, the ATLAS
and CMS experiments have collected a total integrated luminosity of more than
$\simeq$ 30 fm$^{-1}$ each at $\sqrt{s}$=7 and 8 TeV. Up to this date, all observations
(including the Higgs discovery announced at the end of 2011 and
confirmed in the summer/fall of 2012 and spring 2013) are in agreement
with the predictions of the Standard Model, implying limits on the
masses and couplings of exotic particles that are getting more and
more severe~\cite{conferences}.

The main effect of the LHC data on supersymmetry has been to exclude most of the
parameter space corresponding to the more predictive (and falsifiable)
scenarios, such as minimal supergravity (mSUGRA) or the constrained minimal supersymmetric standard model (CMSSM).
However, from the
phenomenological point of view, all SUSY breaking parameters (whose
number, depending on the assumptions, can range between a few to more
than a hundred) are in principle unknown. This implies that the
parameter space of supersymmetry can easily encompass situations
beyond the sensitivity of LHC searches, even when some of the SUSY
particles are light, including the case when the neutralino $\chi$ is
the Lightest Supersymmetric Particle (LSP) and is almost massless~\cite{dreiner}.

In parallel to the LHC, several direct detection experiments searching
for dark matter (DAMA~\cite{dama}, CoGeNT~\cite{cogent},
CRESST~\cite{cresst}, CDMS~\cite{cdms2013}) have recently claimed
possible excesses in their counting rates, which might be explained by
the scattering of a Weakly Interacting Massive Particle (WIMP) with
mass of the order of 10 GeV, and a coherent (scalar) cross section off
nucleons of the order of 10$^{-40}$~cm$^2$.  While these observations
have been challenged by negative results by other experiments such as
XENON100~\cite{xenon100} and CDMS~\cite{cdms}, the robustness of these
constraints has been questioned~\cite{collar,frandsen,hooper},
especially for the lowest range of the WIMP mass.  Since the
neutralino is the most popular explicit realization of a WIMP, the
question on whether supersymmetry can provide a scenario compatible to
the latest constraints from the LHC and capable of explaining the
above results from Dark Matter searches has been discussed in the
literature~\cite{bottino,arbey}.

In one such scenario~\cite{arbey}, in which the SUSY soft masses and
couplings are assumed to be free parameters at the electroweak scale,
the neutralino has a mass of order 10 GeV, is almost degenerate with
the lightest sbottom $\tilde{b}$, and may explain the DAMA, CoGeNT and
CRESST results. The authors of Ref.~\cite{arbey} show that, in spite
of the fact that strong constraints are set by accelerator searches on
light squark masses, when the lightest sbottom mass eigenstate
$\tilde{b}$ is mostly right-handed, it decouples from the Z boson and
goes undetected at LEP. Moreover, when the mass splitting
$m_{\tilde{b}}-m_\chi$ between the sbottom and the neutralino is
smaller than the bottom mass, the sbottom decay $\tilde{b}\rightarrow
b \chi$ is kinematically forbidden, while the decay channel
$\tilde{b}_1\rightarrow \chi s$ is suppressed by the
Cabibbo-Kobayashi-Maskawa (CKM) coupling, possibly increasing the
$\tilde{b}_1$ lifetime up to a value comparable to that of the $b$
hadrons, and preventing a signal from being detected at the LHC by
searches specifically targeted to light
sbottoms~\cite{lhc_sbottom}. The decay rate of the Higgs boson to such
light, invisible sbottom particles can in principle be low enough to
be compatible with the present experimental
data~\cite{higgs_to_sbottom}\footnote{A possible constrain to this
  scenario not discussed in Ref.~\protect\cite{arbey} may arise from
  the modifications introduced by a light colored particle to the
  low--energy running of the strong coupling constant $\alpha_s$
  \protect\cite{bound_alpha_running}. However, we have explicitly
  checked that when the gluino is heavy, the $\alpha_s$ running from
  the $Z$ scale to lower energies is compatible with
  observations.}. Moreover, the small value of $m_{\tilde{b}}-m_\chi$
implies that, in the early Universe, the neutralino coannihilates with
the sbottom, increasing the effective annihilation cross section to
values that drive the predicted thermal relic abundance within the
observational range.

One last bonus of the above scenario is that, as shown for instance in
Figure 2 of Ref.~\cite{arbey}, when $m_{\tilde{b}}-m_\chi$ gets small,
the neutralino--nucleon cross section $\sigma_{\chi N}$ is enhanced,
allowing to reach the range $\sigma_{\chi N}\simeq 10^{-40}$~cm$^2$
needed to explain the possible indications coming from direct
detection experiments. Specifically, the authors of Ref.~\cite{arbey} state in their paper
that they calculate $\sigma_{\chi N}$ using the public code
\mo~\cite{micromegas}.

In the present letter we wish to address the issue of how to calculate
$\sigma_{\chi N}$ in such a specific scenario. For this purpose, we consider only the case in which the lightest sbottom $\tilde{b}$ contributes to the neutralino--nucleon interaction.
In particular, we stress
that it is not valid to introduce a resonance by hand into the tree-level scattering amplitude at $m_{\tilde{b}}=m_\chi+m_b$, as  often made in the existing literature and
available as an option in public codes such as \mo~\cite{micromegas} and
DarkSUSY~\cite{darksusy}.

To show this, we prove that the one-loop neutralino--gluon scattering amplitude, calculated by Drees and Nojiri~\cite{drees_nojiri} (hereafter DN) and Hisano, Ishiwata
and Nagata~\cite{hisano_cross_section} (hereafter HIN), is regular at $m_{\tilde{b}}=m_\chi+m_b$, while it has a pole at $m_{\tilde{b}}=m_\chi-m_b$. Mathematically, the existence of only one pole in the DN and HIN amplitudes, which are computed at zero gluon momentum, is related by Cutkosky rules to the cut in the neutralino self--energy due to the $\chi\to b\tilde{b}$ decay in the region $m_\chi > m_{\tilde{b}}$. Physically, the pole at $m_{\tilde{b}}=m_\chi-m_b$ can be
interpreted as the formation of a resonant state in the nucleon,
specifically either a $C_8 qqq$ R-hadron~\cite{r_hadrons}, with
$C_8$ a $\bar{b}\tilde{b}$ or $b\tilde{b}^*$ color--octet state and $qqq$ the valence quarks of
the nucleon, or a $C_1 qqq$ state, with $C_1$ a $\bar{b}\tilde{b}$ or $b\tilde{b}^*$ color--singlet state. This resonance is of course not kinematically accessible if the neutralino is the LSP.

Since the neutralino-gluon scattering amplitude has no pole in the physical region $m_{\tilde{b}}=m_\chi+m_b$, the common practice of estimating the cross section by
the substitution $m_{\tilde{b}}^{-4}\rightarrow
[(m_{\chi}+m_b)^2-m_{\tilde{b}}^2]^{-2}$ in the propagator of an
effective four--fermion interaction (also including higher--twist
effects) should be discouraged in the case of scattering off bottom
quarks, since it corresponds to adding a spurious pole at $ m_{\tilde{b}}=
m_{\chi}+m_b$, where there is no physical
resonance.
%On the other hand, the prescription $m_{\tilde{b}}^{-4}\rightarrow [(m_{\chi}-m_b)^2-m_{\tilde{b}}^2]^{-2}$ does not show any divergence in the physical domain $m_{\tilde{b}}>m_{\chi}$, and we discuss if it is a good approximation {\bf is it a good approximation?}.
Our considerations can be extended from the specific case of supersymmetry to other similar
cases in which the Dark Matter particle scatters off nucleons through the
exchange of a $b$--flavored state almost degenerate in mass with the dark matter particle, such as,
for instance,  in theories with extra space-time dimensions~\cite{ued} and in the ``mass--degenerate
dark matter'' scenarios of Refs.~\cite{vogl} and~\cite{hisano}.

We divide this article in several sections. In Section
\ref{sec:cross_section} we present the general ingredients entering a calculation of the neutralino--nucleon cross
section, and introduce some notation to guide the discussion. In the following Sections (\ref{sec:heavysquark} to \ref{sec:darksusy}), we review several calculations existing in the literature: the heavy squark limit, the pole prescription we question, the one-loop results of DN and HIN, and the options currently available in the pubic codes \mo\ and DarkSUSY.  In Section \ref{sec:results} we quantitatively
discuss the behavior of the cross section comparing the various methods we review for the specific case of near degeneracy between the neutralino and the
sbottom. We finally collect some useful formulas in the Appendix.

\section{Generalities}
\label{sec:cross_section}

In this paper, we focus on the contribution to the spin--independent (scalar) neutralino--nucleon cross section generated by the neutralino--quark--squark interaction Lagrangian:
\begin{align}
\mathcal{L}_{\tilde{q}q\chi} = \tilde{q} \,\, \overline{q}  \, \left( \aq + \bq \gamma_5\right)  \chi + {\rm h.c.}
\label{eq:lagrangian}
\end{align}
In particular, we neglect all other contributions to the cross
section, such as those coming from Higgs exchange and $Z$--boson exchange (the
latter contributes only to the spin--dependent cross section). In Eq.~(\ref{eq:lagrangian}), $\chi$, $q$, and $\tilde{q}$ are the neutralino, quark, and squark fields, and the coupling constants $\aq$ and $\bq$ are functions of the model parameters. In particular, in our numerical analysis, we are interested in the bottom quark and the lightest sbottom squark.

An effective Lagrangian $\mathcal{L}_{\chi N}^{\rm SI}$ is defined to describe neutralino-nucleon spin-independent (SI) scattering at zero momentum transfer. It can be written in terms of an effective neutralino-nucleon coupling $f$ (introduced by DN) as
\begin{align}
\mathcal{L}_{\chi N}^{\rm SI} = f \, \overline{\chi} \, \chi  \, \overline{N} N,
\label{eq:f_psi_psibar}
\end{align}
where $N$ is the Dirac field of the nucleon. Notice that since the neutralino is a Majorana particle, this formula implies that the four-particle $\chi\chi NN$ vertex in the Feynman rules is $2f$.

The cross section $\sigma_{\chi N} $ for the non-relativistic elastic scattering of a neutralino of mass $m_\chi$ off a nucleon of mass $m_N$ then follows as
\begin{align}
\sigma_{\chi N} = \frac{4m_\chi^2m_N^2}{\pi(m_\chi+m_N)^2} \, f^2.
\label{eq:cross_section}
\end{align}

The fundamental Lagrangian contains Standard Model interactions plus the interaction $\mathcal{L}_{\tilde{q}q\chi}$ in Eq.~(\ref{eq:lagrangian}) between squarks, quarks, and neutralinos. Each quark field with mass much higher than the QCD scale $\Lambda_{\rm QCD} \sim 400$ MeV ($Q=c,b,t$) can be integrated out of the theory and its interactions replaced by terms containing effective operators involving gluons.
Thus the effective SI neutralino-parton Lagrangian reads
\begin{align}
\mathcal{L}_{\rm eff}^{\rm SI} & =  \sum_{q=u,d,s}  f_q \, m_q \, \overline{q} \, q \, \overline{\chi} \, \chi  + \sum_{Q=c,b,t} f^{(Q)}_G \, G^{a\mu\nu} G^{a}_{\mu\nu} \, \overline{\chi} \, \chi
\nonumber \\ & +  \sum_{q=u,d,s}  \left( g^{(1)}_q \, \frac{\overline{\chi} i \partial^\mu \gamma^\nu \chi}{m_\chi} + g^{(2)}_q \, \frac{\overline{\chi} i \partial^\mu i \partial^\nu \chi}{m_\chi^2}   \right) \mathcal{O}^{(2)}_{q\mu\nu}
\nonumber \\ & + \sum_{Q=c,b,t}  \left( g^{(1,Q)}_G \, \frac{\overline{\chi} i \partial^\mu \gamma^\nu \chi}{m_\chi} + g^{(2,Q)}_G \, \frac{\overline{\chi} i \partial^\mu i \partial^\nu \chi}{m_\chi^2} \right) \mathcal{O}^{(2)}_{G\mu\nu}  .
\label{eq:Lqg}
\end{align}
Here $G^{a}_{\mu\nu}$ is the gluon field strength, while $\mathcal{O}^{(2)}_{q\mu\nu}$ and $\mathcal{O}^{(2)}_{G\mu\nu}$ are the quark and gluon twist-2 operators
\begin{align}
\mathcal{O}^{(2)}_{q\mu\nu} & = \frac{i}{2} [ \overline{q} \gamma_\mu \partial_\nu q + \overline{q} \gamma_\nu \partial_\mu q - \frac{1}{2} \overline{q} \gamma^\alpha \partial_\alpha q g_{\mu\nu} ],
\\
\mathcal{O}^{(2)}_{G\mu\nu} & = G^{a}_{\phantom{a}\mu\rho} G^{a\rho}_{\phantom{a\rho}\nu} + \frac{1}{4} g^{\mu\nu} G^{a\alpha\beta} G^{a}_{\alpha\beta} .
\end{align}
The twist-2 operators are symmetric and are the traceless parts of the energy--momentum tensors,
\begin{align}
T_{q\mu\nu} & = \frac{1}{4} g_{\mu\nu} m_q \overline{q} q + \mathcal{O}^{(2)}_{q\mu\nu} ,
\\
T_{g\mu\nu} & = \frac{1}{4} g_{\mu\nu} G^{a\alpha\beta} G^{a}_{\alpha\beta} + \mathcal{O}^{(2)}_{G\mu\nu} .
\end{align}
Our definitions of $f_q$, $g^{(1)}_q$, and $g^{(2)}_q$ coincide with those of HIN, while HIN's $f_G$ and $g^{(i)}_G$ are
\begin{align}
f_G & = \sum_{Q=c,b,t} f^{(Q)}_G,
\\
g^{(i)}_G & = \sum_{Q=c,b,t} g^{(i,Q)}_G  \qquad \text{$(i=1,2)$}.
\end{align}

The coefficients $f^{(Q)}_G$, $f_q$, $g_G^{(i,Q)}$, and $g_q^{(i)}$
($i=1,2$) are fixed by matching $\mathcal{L}^{\rm SI}_{\rm eff}$ to
the zero-momentum transfer limit of suitable diagrams computed using
the fundamental theory. For a plane--wave neutralino $\chi$ of momentum
$p^\mu$ one has
\begin{align}
\frac{\overline{\chi} i \partial^\mu \gamma^\nu \chi}{m_\chi} & = \frac{\overline{\chi} i \partial^\mu i \partial^\nu \chi}{m_\chi^2} = \frac{p^\mu p^\nu}{m_\chi^2} \overline{\chi} \chi.
\label{eq:chiexpectation}
\end{align}
Thus only the sums
\begin{align}
g^{(1)}_G + g^{(2)}_G,
\qquad
g^{(1)}_q + g^{(2)}_q ,
\end{align}
enter the neutralino--nucleon effective Lagrangian.

The quark trace coefficient $f_q$ and the quark twist-2 coefficients
$g^{(i)}_q$ are in principle obtained by taking the zero-momentum
transfer limit of the $\chi\chi q q$ diagrams in
Fig.~\ref{fig:dia1}. However this calculation can be explicitly
carried out only in the limit of heavy squarks (see DN and
HIN). Formally, one matches the forward amplitudes of $\chi q \to \chi
q$ obtained with the fundamental and effective Lagrangian.  For a
plane--wave neutralino $|\chi\rangle$ of momentum $p^\mu$, and a
plane-wave quark $|q\rangle$ of momentum $k^\mu$, we find
\begin{align}
\Big\langle \chi q \Big| \mathcal{L}_{\rm eff}^{\rm SI} \Big| \chi q \Big\rangle
 & = \Bigg[ f_q + \Bigg( g^{(1)}_q + g^{(2)}_q \Bigg) \Bigg( \frac{(p\cdot k)^2}{m_\chi^2m_q^2} - \frac{1}{4} \Bigg) \Bigg] \Big\langle \chi q \Big| m_q \overline{q}q\overline{\chi}\chi \Big| \chi q \Big\rangle .
\label{eq:Leffchiq}
\end{align}

\begin{figure}[t]
\begin{center}
\includegraphics[width=0.80\columnwidth]{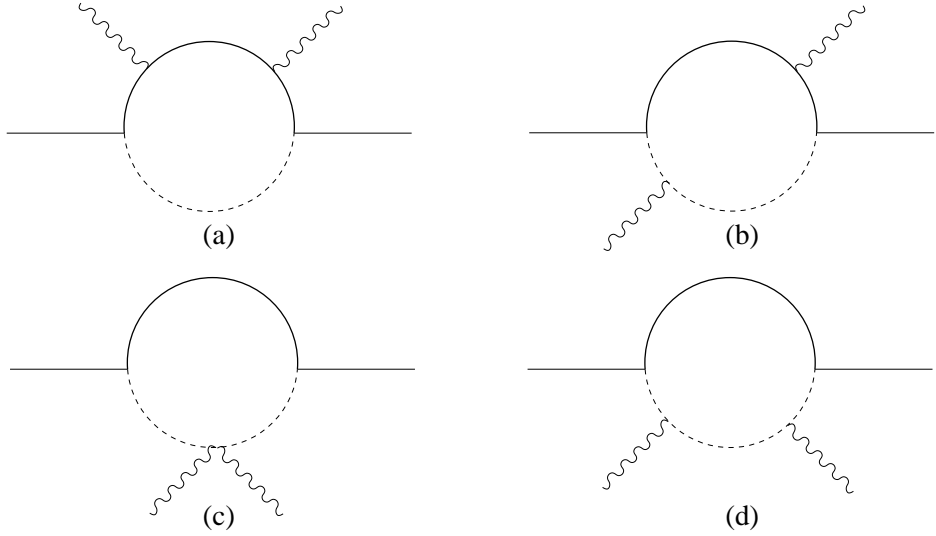}
\end{center}
\caption{Diagrams contributing to the neutralino--gluon effective
  Lagrangian via squark exchange. Neutralinos are shown with thin
  solid lines, quarks with thick solid lines, squarks with dashed
  lines and gluons with wavy lines. Diagrams with exchanged gluons
  should be added. }
\label{fig:dia2}
\end{figure}

\begin{figure}[t]
\begin{center}
\includegraphics[width=0.3\columnwidth]{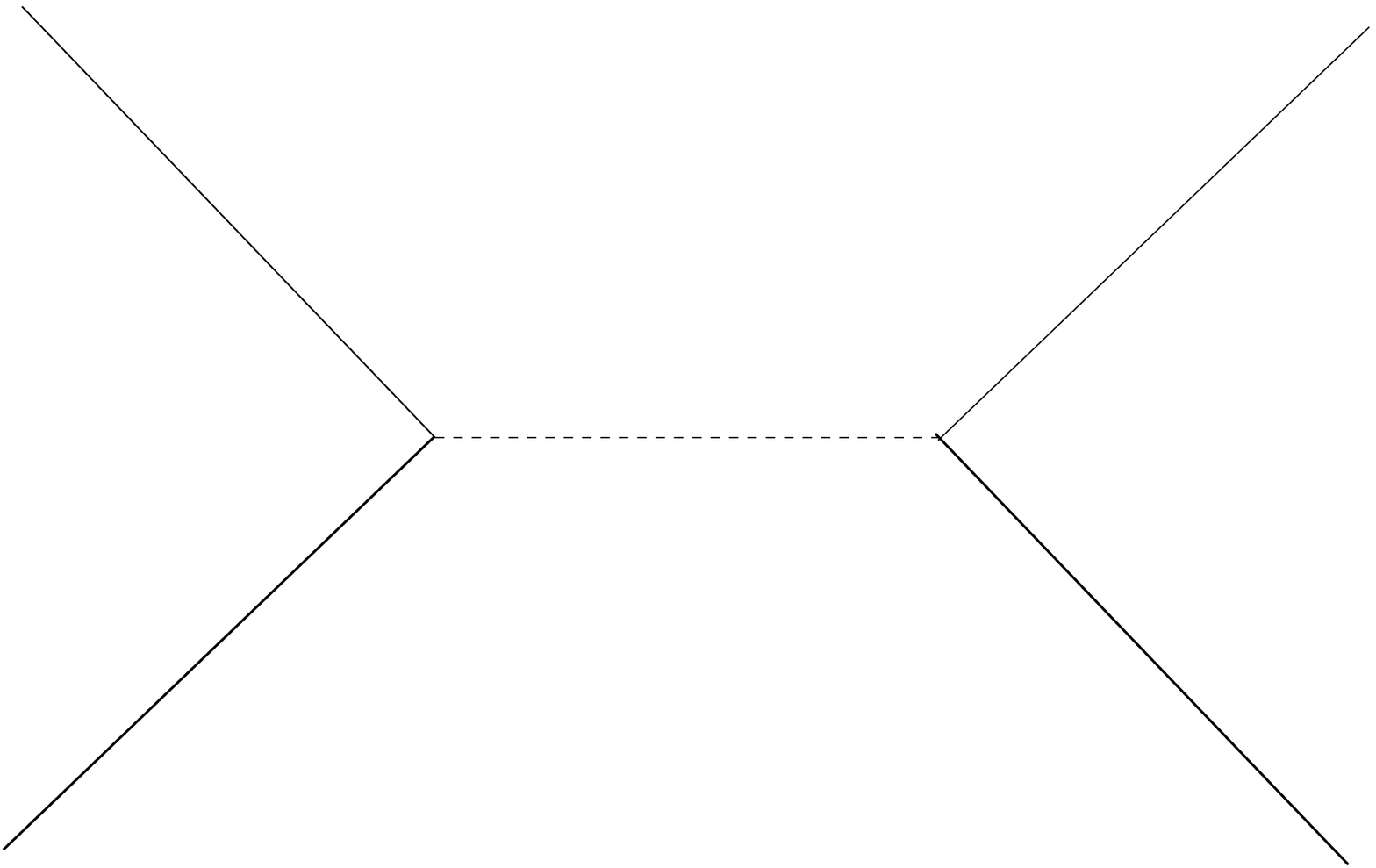}\hspace{6em}
\includegraphics[width=0.3\columnwidth]{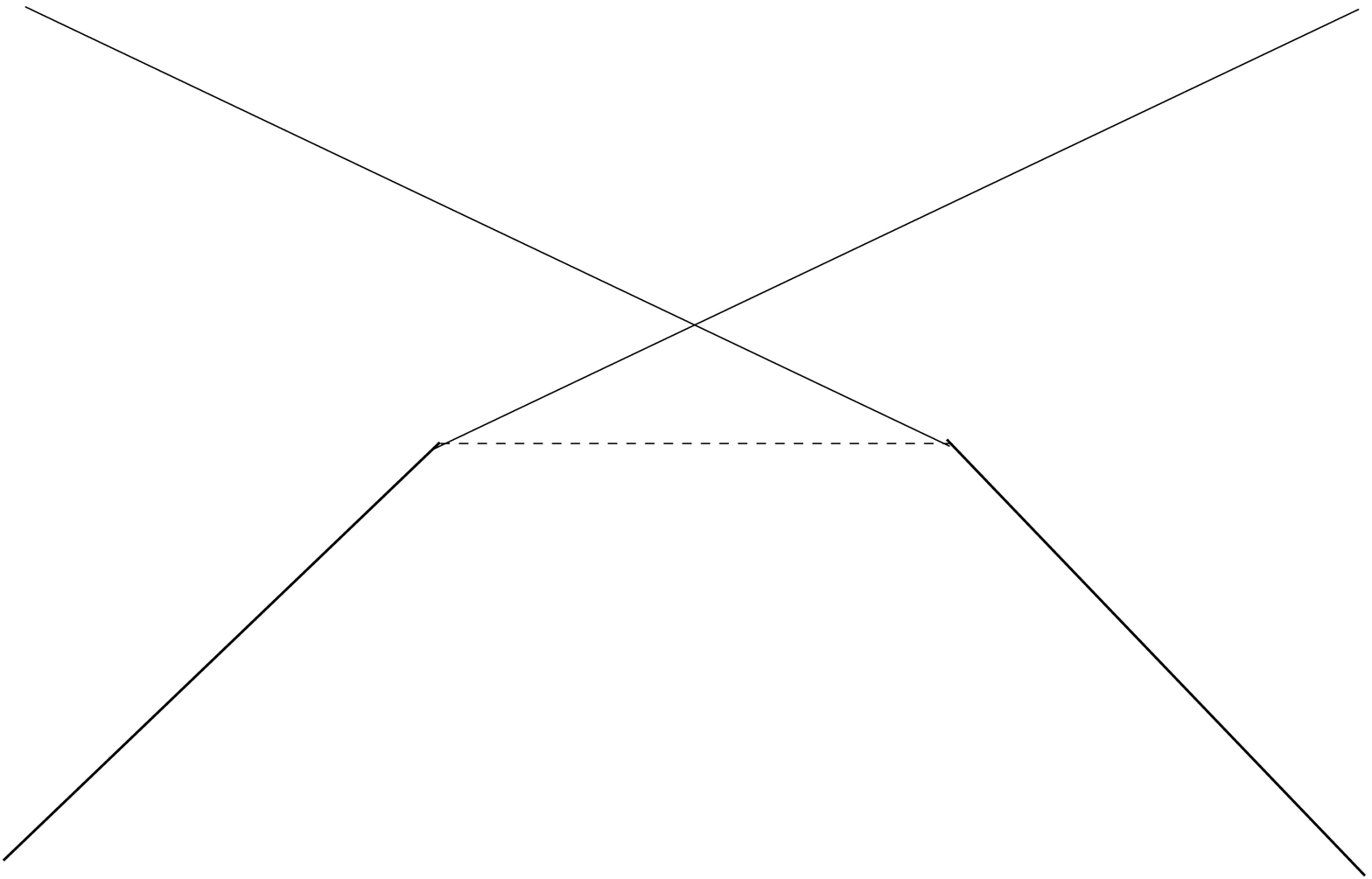}
\end{center}
\caption{Diagrams contributing to the neutralino--quark effective
  Lagrangian via squark exchange. Line--style conventions are the same
  as in Fig.\protect\ref{fig:dia2}. }
\label{fig:dia1}
\end{figure}

The gluon trace coefficients $f^{(Q)}_G$ and the gluon twist-2 coefficients $g^{(i,Q)}_G$ are obtained by taking the zero-momentum transfer limit of the $\chi\chi g g$ diagrams in Fig.~\ref{fig:dia2} (this is what ``integrating out heavy quarks'' means). They have been computed to one-loop by DN (and in some cases to two-loops by HIN). It is clear that each quark flavor $q$ contributes in principle a term to $f_G$ and $g^{(i)}_G$. However, the light quarks $u$, $d$, $s$ cannot to be included as quark loops, since such a one-loop QCD calculation at large distances (small loop momenta of order $m_q$) would not be a good perturbative approximation.

We stress that each quark flavor is to be included in the effective Lagrangian $\mathcal{L}^{\rm SI}_{\rm eff}$ either in the quark terms (if not integrated out) or in the gluon terms (if integrated out), but not in both. One can either include heavy quarks in the form
\begin{align}
\mathcal{L}_{\chi Q} = f_Q \, m_q \, \overline{Q} \, Q \, \overline{\chi} \, \chi  +  \left( g^{(1)}_Q \, \frac{\overline{\chi} i \partial^\mu \gamma^\nu \chi}{m_\chi} + g^{(2)}_Q \, \frac{\overline{\chi} i \partial^\mu i \partial^\nu \chi}{m_\chi^2}   \right) \mathcal{O}^{(2)}_{Q\mu\nu},
\label{eq:LQ1}
\end{align}
where the index $Q$ now refers to a heavy quark, or integrate out heavy quarks and include them in the form
\begin{align}
\mathcal{L}_{\chi G}^{(Q)} = f^{(Q)}_G \, G^{a\mu\nu} G^{a}_{\mu\nu} \, \overline{\chi} \, \chi +
 \left( g^{(1,Q)}_G \, \frac{\overline{\chi} i \partial^\mu \gamma^\nu \chi}{m_\chi} + g^{(2,Q)}_G \, \frac{\overline{\chi} i \partial^\mu i \partial^\nu \chi}{m_\chi^2} \right) \mathcal{O}^{(2)}_{G\mu\nu} .
\end{align}
In $\mathcal{L}_{\chi Q}$, the heavy quark trace operator $m_Q \overline{Q} Q$ may be rewritten in terms of the gluon trace operator $G^{a\mu\nu} G^{a}_{\mu\nu}$ using the operator heavy-quark relation~\cite{svz}
\begin{align}
m_Q \overline{Q} Q = - \frac{\alpha_s}{12\pi} G^{a\mu\nu} G^{a}_{\mu\nu} .
\label{eq:svz}
\end{align}
This leads to
\begin{align}
\mathcal{L}_{\chi Q} = - \frac{12\pi}{\alpha_s} f_Q \, G^{a\mu\nu} G^{a}_{\mu\nu} \, \overline{\chi} \, \chi +  \left( g^{(1)}_Q \, \frac{\overline{\chi} i \partial^\mu \gamma^\nu \chi}{m_\chi} + g^{(2)}_Q \, \frac{\overline{\chi} i \partial^\mu i \partial^\nu \chi}{m_\chi^2}   \right) \mathcal{O}^{(2)}_{Q\mu\nu}.
\end{align}
This equivalent form of $\mathcal{L}_{\chi Q}$ helps in computing matrix elements, but the heavy quarks have actually not been integrated out and the coefficients $f_Q$ and $g^{(i)}_Q$ remain those of the tree-level neutralino--quark interaction.
Some authors use a hybrid form
\begin{align}
f^{(Q)}_G \, G^{a\mu\nu} G^{a}_{\mu\nu} \, \overline{\chi} \, \chi +
  \left( g^{(1)}_Q \, \frac{\overline{\chi} i \partial^\mu \gamma^\nu \chi}{m_\chi} + g^{(2)}_Q \, \frac{\overline{\chi} i \partial^\mu i \partial^\nu \chi}{m_\chi^2}   \right) \mathcal{O}^{(2)}_{Q\mu\nu},
 \label{eq:hybrid}
\end{align}
where the first term is computed using the neutralino--gluon loop diagrams in Fig.~\ref{fig:dia2}, and the second term is computed using the tree-level neutralino--quark diagrams in Fig.~\ref{fig:dia1}. These approaches are not generally equivalent. In particular, the $f_Q$ and $g^{(i)}_Q$ coefficients are only calculable in the limit of heavy squark masses $m_{\tilde{Q}} \gg m_\chi, m_Q$. When in the literature they are extrapolated to finite squark masses, a spurious propagator pole at $m_{\tilde{Q}} = m_\chi + m_Q$ is often introduced, which is absent in the corresponding coefficients $f^{(Q)}_G$ and $g^{(i,Q)}_G$. Since we are interested in this regime, and the hybrid form above is questionable, we integrate out all heavy quarks, i.e.\ we use the form $\mathcal{L}_{\chi G}^{(Q)}$ for $Q=c,b,t$.

For the Lagrangian in Eq.~(\ref{eq:lagrangian}), DN's matching of the one-loop $\chi g \to \chi g$ amplitude in the fundamental and effective theories gives
\begin{align}
-\frac{12\pi}{\alpha_s} f^{(Q)}_{G} & = \frac{ \aQ^2-\bQ^2 }{4} m_Q f_{D}^{(Q)}
+ \frac{ \aQ^2+\bQ^2}{4} m_\chi f_{S}^{(Q)} ,
\label{eq:ourfG}
\\
g^{(1,Q)}_{G} + g^{(2,Q)}_{G} & = \frac{ \aQ^2-\bQ^2 }{4} m_Q g_{D}^{(Q)}
+ \frac{ \aQ^2+\bQ^2}{4} m_\chi g_{D}^{(Q)} .
\label{eq:ourgG}
\end{align}
Here $f_{D}^{(Q)}$, $f_{S}^{(Q)}$, $g_{D}^{(Q)}$ and $g_{S}^{(Q)}$ are expressed in terms of the DN loop integrals $I_n(m_{\tilde{Q}},m_Q,m_\chi)$ by the relations
\begin{align}
f_{D}^{(Q)} & = m_\chi^2 I_3 - \frac{3}{2} I_1,
\label{eq:fpmgpm1}
\\
f_{S}^{(Q)} & = m_\chi^2 I_4 + \frac{1}{2} I_5  - \frac{3}{2} I_2 ,
\label{eq:fpmgpm2}
\\
g_{D}^{(Q)} & = \frac{\alpha_s}{3\pi} m_\chi^2 I_3.
\label{eq:fpmgpm3}
\\
g_{S}^{(Q)} & = \frac{\alpha_s}{3\pi} \left( m_\chi^2 I_4 + \frac{1}{2} I_5 \right).
\label{eq:fpmgpm4}
\end{align}
The expressions of the loop integrals $I_n$ defined by DN are provided for completeness in the Appendix, where we also give their analytic continuation for $m_\chi < m_{\tilde{q}}$.

%%%%%%%%%%%%%%%%%%%%%%%%%%%%%%%%%%%%%%%%%%%%%%%%%%%%%%%
%%%%%%%%%%%%%%%%%%%%%%%%%%%%%%%%%%%%%%%%%%%%%%%%%%%%%%%

%%%%%%%%%%%%%%%%%%%%%%%%%%%%%%%%%%%%%%%%%%%%%%%%%%%%%%%
%%%%%%%%%%%%%%%%%%%%%%%%%%%%%%%%%%%%%%%%%%%%%%%%%%%%%%%

The final expression of $f$ is found by taking nucleonic matrix elements of the quark and gluon operators in $\mathcal{L}_{\rm eff}^{\rm SI}$. These matrix elements must be obtained experimentally, and are traditionally parametrized in terms of the quantities $f_{Tq}$, $f_{TG}$, $q(2,\mu^2)+\overline{q}(2,\mu^2)$, and $G(2,\mu^2)$ as
\begin{align}
\langle N | m_q \overline{q} q | N \rangle & = m_N f_{Tq} \, \langle N | \overline{N} N | N \rangle ,
\\
\Big\langle N \Big| - \frac{\alpha_s}{12\pi}G^{a\mu\nu} G^{a}_{\mu\nu} \Big| N \Big\rangle & = \frac{2}{27} m_N f_{TG} \, \langle N | \overline{N} N | N \rangle ,
\label{eq:GGmatrixel}
\\
\big\langle N \big| \mathcal{O}^{(2)}_{q\mu\nu} \big| N \big\rangle & = \frac{1}{m_N} \left( p_{N\mu} p_{N\nu} - \tfrac{1}{4} m_N^2 g_{\mu\nu} \right) \Big( q(2,\mu^2) + \overline{q}(2,\mu^2) \Big) \, \langle N | \overline{N} N | N \rangle ,
\\
\big\langle N \big| \mathcal{O}^{(2)}_{G\mu\nu} \big| N \big\rangle & = \frac{1}{m_N} \left( p_{N\mu} p_{N\nu} - \tfrac{1}{4} m_N^2 g_{\mu\nu} \right) G(2,\mu^2) \, \langle N | \overline{N} N | N \rangle .
\end{align}
Physically, $G(2,\mu^2)$, $q(2,\mu^2)$, and $\bar{q}(2,\mu^2)$ are the
second moments of the parton distribution functions for gluons,
quarks, and antiquarks at the renormalization scale $\mu$~\cite{drees_nojiri}:
\begin{align}
q(2,\mu^2) = \int_0^1 dx \, x \, q(x,\mu^2),
\end{align}
and similarly for the others.
The quantities $f_{Tq}$ and $f_{TG}$ are the fractional contributions of quarks and gluons to the mass of the nucleon. The latter property, which also accounts for the factor of -2/27 in Eq.~(\ref{eq:GGmatrixel}), derives from the expression of the trace of the nucleon energy momentum tensor~\cite{Ji:1994av,Ji:1995sv}
\begin{align}
 \langle N |  T^{\mu}_{\phantom{\mu}\mu} | N \rangle = \Big\langle N \Big| -\frac{27}{2} \frac{\alpha_s}{12\pi} G^{a\mu\nu} G^{a}_{\mu\nu} + \sum_{q=u,d,s} m_q \, \overline{q} \, q \Big| N \Big\rangle.
\end{align}
This relation leads to $f_{TG} + f_{Tu} + f_{Td} + f_{Ts} = 1$. The twist-2 operators are also related to the energy-momentum tensor, namely its traceless symmetric part,
\begin{align}
\langle N | T_{\mu\nu} - \frac{1}{4} g_{\mu\nu} T^{\alpha}_{\phantom{\alpha}\alpha} | N \rangle
= \Big\langle N \Big| \mathcal{O}^{(2)}_{G\mu\nu}  + \sum_{q=u,d,s,c,b,t} \mathcal{O}^{(2)}_{q\mu\nu} \Big| N \Big\rangle.
\end{align}
This relation leads to the sum rule
\begin{align}
G(2,\mu^2) + \sum_{q=u,d,s,c,b,t} \Big( q(2,\mu^2) + \overline{q}(2,\mu^2) \Big) = 1,
\end{align}

In practice, one obtains $f_{Tu}$, $f_{Td}$, $f_{Ts}$ for the light quarks (and $f_{TG}$ for the gluons) from chiral perturbation theory and the pion--nucleon sigma-term or from lattice QCD. And one may also obtain $f_{Tc}$, $f_{Tb}$, and $f_{Tt}$ for the heavy quarks from a heavy quark expansion, which for a heavy quark $Q$ gives the relation~\cite{svz}
\begin{align}
f_{Tc}=f_{Tb}=f_{Tt} = \frac{\langle N | m_Q \overline{Q} Q | N \rangle}{\langle N | m_N \overline{N} N | N \rangle} =  \frac{\langle N | - (\alpha_s/12\pi) G^{a\mu\nu} G^{a}_{\mu\nu} | N \rangle}{\langle N | m_N \overline{N} N | N \rangle} = \frac{2}{27} f_{TG} .
\label{eq:qqbar_heavy}
\end{align}
In the numerical analysis in Section~\ref{sec:results} we take the numerical values of $f_{Tq}$, $f_{TG}$, $q(2,\mu^2)+\overline{q}(2,\mu^2)$, and $G(2,\mu^2)$ from HIN.

Finally, we obtain the general formula for the effective neutralino-nucleon coupling $f$,
\begin{align}
f & = m_N \left[ \frac{2}{27} f_{TG} \sum_{Q=c,b,t} \left( -\frac{12\pi}{\alpha_s} f^{(Q)}_G \right)  + \sum_{q=u,d,s} f_q f_{Tq} \right]
\nonumber \\
& +
\frac{3}{4} m_N \Bigg[ G(2,\mu^2) \sum_{Q=c,b,t} [ g^{(1,Q)}_{G}+g^{(2,Q)}_{G} ] \,\,\, + \sum_{q=u,d,s} [g^{(1)}_q+g^{(2)}_q] \, \Big( q(2,\mu^2) + \overline{q}(2,\mu^2) \Big)   \Bigg]
.
\label{eq:f}
\end{align}
In particular, our expression for the contribution of the bottom quark we are specifically interested in is, separating the $a_b^2-b_b^2$ and $a_b^2+b_b^2$ parts,
\begin{align}
\left. f \right|_b = m_N \Bigg\{ &  \frac{ a_b^2-b_b^2 }{4} m_b \left[ \frac{2}{27} \, f_{TG}\, f_{D}^{(Q)} + \frac{3}{4} \,G(2,\mu^2) \, g_{D}^{(Q)} \right]
\nonumber \\ &
+ \frac{ a_b^2+b_b^2}{4} m_\chi \left[ \frac{2}{27} \, f_{TG}\, f_{S}^{(Q)} + \frac{3}{4} \,G(2,\mu^2) \, g_{S}^{(Q)}  \right] \Bigg\} .
\label{eq:fb}
\end{align}
The loop integrals $f^{(Q)}_{S,D}$ and $g^{(Q)}_{S,D}$ are given in Eqs.~(\ref{eq:fpmgpm1})-(\ref{eq:fpmgpm4}).

In the following sections we compare our formula for $f|_Q$ with those that have appeared in the literature, reviewing the various expressions and approximations for the coefficients $f^{(Q)}_G$, $f_q$, $g^{(i,Q)}_G$, $g^{(i)}_{q}$.

\section{Heavy squark limit}
\label{sec:heavysquark}

The heavy squark limit $m_{\tilde{q}}\gg m_{\chi},m_q$ was presented
very early in the literature
\cite{Goodman:1984dc,Gaisser:1986ha,griest,gg91,ellis}. In these early
papers, the propagator in the tree-level diagrams of
Fig.~\ref{fig:dia1} is contracted to a point, and the gluon diagrams
in Fig.~\ref{fig:dia2} are not included. In the heavy squark limit
(HSL), the effective Lagrangian coefficients \coeffs are
\begin{align}
f^{\rm HSL}_q  & = - \frac{\aq ^2-\bq^2}{4m_qm_{\tilde{q}}^2}  , & (q=u,d,s)
\label{eq:HSL1}
\\
-\frac{12\pi}{\alpha_s} f_G^{(Q)\rm HSL} & = - \frac{\aQ^2-\bQ^2}{4m_Qm_{\tilde{Q}}^2}  , & (Q=c,b,t)
\label{eq:HSL2}
\\
g^{(i)\rm HSL}_q & = g^{(i)\rm HSL}_G = 0 .
\label{eq:HSL3}
\end{align}
For the lightest sbottom case of interest to us, the heavy squark limit expression for $f$ is
\begin{align}
\left. f \right|_b^{\rm HSL} = - m_N \frac{2}{27} f_{TG} \frac{a_b^2-b_b^2}{4m_bm_{\tilde{b}}^2} .
\label{eq:heavy_squark}
\end{align}
This coincides with the heavy squark limit of our Eq.~(\ref{eq:fb}) to
order $m_{\tilde{b}}^{-2}$.  Notice that the previous expression
vanishes for $\aq^2=\bq^2$.

\section{Drees and Nojiri}
\label{sec:DN}

Drees and Nojiri~\cite{drees_nojiri} write the neutralino--quark effective Lagrangian as
\begin{align}
\mathcal{L}^{\rm SI}_{q,\rm eff} & = \hat{f}_q \overline{\chi} \chi \overline{q} q + \hat{g}_q \overline{\chi} \gamma^\mu \partial^\nu \chi (\overline{q} \gamma_\mu \partial_\nu q - \partial_\nu \overline{q} \gamma_\mu q )
\\ & = \Big( \hat{f}_q - \frac{1}{2} \hat{g}_q m_q m_\chi\Big) \overline{\chi} \chi \overline{q} q  - 2 \hat{g}_q \overline{\chi} i \gamma^\mu \partial^\nu \chi  \mathcal{O}^{(2)}_{q\mu\nu} .
\end{align}
In this equation, $\hat{f}_q$ and $\hat{g}_q$ denote the coefficients called $f_q$ and $g_q$ in DN. Keeping only the squark terms in the DN coefficients $\hat{f}_q$ and $\hat{g}_q$, we read off the $f_q$ and $g^{(1)}_q+g^{(2)}_q$ coefficients of DN,
\begin{align}
f^{\rm DN}_q & = - \frac{1}{4m_q} \frac{\aq ^2-\bq^2}{m_{\tilde{q}}^2 - (m_\chi+m_q)^2} + \frac{m_\chi}{8} \frac{\aq ^2+\bq^2}{[m_{\tilde{q}}^2 - (m_\chi+m_q)^2]^2} ,
\label{eq:dnq}
\\
g^{(1)\rm DN}_q + g^{(2)\rm DN}_q & = \frac{m_\chi}{2} \frac{\aq ^2+\bq^2}{[m_{\tilde{q}}^2 - (m_\chi+m_q)^2]^2} .
\label{eq:dnq2}
\end{align}
Notice that the quark-mass dependence of the squark propagator in
the expressions above is not explicitly derived in DN, and is in fact different from the expressions in other sections.

The coefficients $f_G^{(Q)}$ and $g^{(i,Q)}_G$ for squark exchange can be read off the neutralino--gluon effective Lagrangian in DN (their Eqs.~(17) and (19)),
\begin{align}
\mathcal{L}_{G,\rm eff}^{\rm SI} & = (B_D+B_S)\overline{\chi} \chi G^{a\mu\nu} G^{a}_{\mu\nu} -(B_{1D}+B_{1S}) \overline{\chi} \partial_\mu\partial_\nu \chi G^{a\mu\rho} G^{a\phantom{\rho}\nu}_{\phantom{a}\rho}
\nonumber \\ & + B_{2S} \overline{\chi} ( i \partial_\mu\gamma_\nu+i\partial_\nu\gamma_\mu) \chi  G^{a\mu\rho} G^{a\phantom{\rho}\nu}_{\phantom{a}\rho} \nonumber \\
& =  [ B_D+B_S  - \tfrac{1}{4} m_\chi^2 (B_{1D}+B_{1S})  - \tfrac{1}{2} m_\chi B_{2S} ] \overline{\chi} \chi G^{a\mu\nu} G^{a}_{\mu\nu}
\nonumber \\ & \quad
 +  [ 2 B_{2S} \overline{\chi} i \partial^\mu\gamma^\nu \chi + (B_{1D}+B_{1S}) \overline{\chi} i\partial^\mu i \partial^\nu \chi  ] \mathcal{O}^{(2)}_{G\mu\nu}  .
\label{eq:DN}
\end{align}
Here
\begin{align}
B_D & = \frac{\alpha_s}{4\pi} \frac{1}{8} \sum_q (a_{\tilde{q}}^2-b_{\tilde{q}}^2) m_q I_1(m_{\tilde{q}},m_q,m_\chi) ,\nonumber \\
B_S & = \frac{\alpha_s}{4\pi} \frac{1}{8} \sum_q (a_{\tilde{q}}^2+b_{\tilde{q}}^2) m_\chi I_2(m_{\tilde{q}},m_q,m_\chi) ,\nonumber \\
B_{1D} & = \frac{\alpha_s}{4\pi} \frac{1}{3} \sum_q (a_{\tilde{q}}^2-b_{\tilde{q}}^2) m_q I_3(m_{\tilde{q}},m_q,m_\chi) ,\nonumber \\
B_{1S} & = \frac{\alpha_s}{4\pi} \frac{1}{3} \sum_q (a_{\tilde{q}}^2+b_{\tilde{q}}^2) m_\chi I_4(m_{\tilde{q}},m_q,m_\chi) , \nonumber \\
B_{2S} & = \frac{\alpha_s}{4\pi} \frac{1}{12} \sum_q (a_{\tilde{q}}^2+b_{\tilde{q}}^2) I_5(m_{\tilde{q}},m_q,m_\chi) .
\label{b_factors}
\end{align}

From Eq.~(\ref{eq:DN}) we extract the coefficients
\begin{align}
-\frac{12\pi}{\alpha_s} f^{(Q)\rm DN}_{G} & = \frac{ \aQ^2-\bQ^2 }{4} m_Q \left(m_\chi^2 I_3  - \frac{3}{2} I_1 \right)
+ \frac{ \aQ^2+\bQ^2}{4} m_\chi \left( m_\chi^2 I_4 + \frac{1}{2} I_5 - \frac{3}{2} I_2  \right),
\label{eq:dnfG}
\\
g^{(1,Q)\rm DN}_{G} + g^{(2,Q)\rm DN}_{G} & = \frac{\alpha_s}{3\pi}  \left[ \frac{ \aQ^2-\bQ^2 }{4} m_Q m_\chi^2 I_3 + \frac{ \aQ^2+\bQ^2 }{4} \left(  m_\chi^2 I_4 + \frac{1}{2} I_5 \right) \right] .
\label{eq:dngG}
\end{align}
In the heavy squark limit ($m_{\tilde{q}} \gg m_\chi, m_q$), one finds agreement with Eqs.~(\ref{eq:HSL2})-(\ref{eq:HSL3}),
\begin{align}
-\frac{12\pi}{\alpha_s} f^{(Q)\rm DN}_{G}& \simeq - \frac{\aQ^2-\bQ^2 }{4m_Q m_{\tilde{Q}}^2} + O\!\left( \frac{1}{m_{\tilde{Q}}^4} \right)
\\
g^{(1,Q)\rm DN}_{G} + g^{(2,Q)\rm DN}_{G}& \simeq  O\!\left( \frac{1}{m_{\tilde{Q}}^4} \right) .
\label{eq:I2_i5}
\end{align}

We point out that our trace terms for $f$ agree with those of DN, but
DN include heavy quarks into the twist-2 gluon and quark terms in
Eq.~(\ref{eq:f}) in a way different from ours. We include all heavy
quarks in both the gluon and quark twist-2 terms, while DN include
only some, according to the following scheme. (a) They do not include
the top quark in the quark twist-2 terms, which is a good
approximation since the top quark PDF in the nucleon is
negligible. (b) They do not include the $c$ and $b$ quarks in the
twist-2 gluon term (see their equation 46). And (c) they include the
bottom quark either in the twist-2 gluon term or in the twist-2 quark
term, according to which gives the smallest contribution. This is
because they consider the possibility of a light sbottom
$m_{\tilde{b}} \sim m_\chi$, for which their $g^{(1)\rm DN}_b +
g^{(2)\rm DN}_b $ coefficient, Eq.~(\ref{eq:dnq2}), diverges.
% (on the contrary, as discussed below, our $g^{(1)}_b + g^{(2)}_b $,
% Eq.~(\ref{eq:g1Q}), does not diverge).

For the lightest sbottom quark $\tilde{b}_1$, DN advocate the following prescription to avoid what they call ``the spurious pole'' in the twist-2 quark coefficients $g^{(1)\rm DN}_{b}+g^{(2)\rm DN}_{b}$ at $m_{\tilde{b}_1}=m_\chi+m_b$. Compute the amplitude $f$ in two separate ways, including the lightest squark $\tilde{b}_1$ either in the twist-2 gluon term or in the twist-2 quark term, and take the amplitude that gives the smallest cross section. In formulas,  with $b_1$ referring to the lightest sbottom contributions and $b_2$ to the heaviest sbottom contributions,
\begin{align}
f^{\rm DN}_{G(2)} & = m_N \left[ \frac{2}{27} f_{TG} \sum_{Q=c,b,t} \left( -\frac{12\pi}{\alpha_s} f^{(Q)\rm DN}_G \right)  + \sum_{q=u,d,s} f^{\rm DN}_q f_{Tq} \right]
\label{eq:dnfb1}
\nonumber \\
& +
\frac{3}{4} m_N \Bigg[ G(2,\mu^2) \sum_{Q=b_1,t} [ g^{(1,Q)\rm DN}_{G}+g^{(2,Q)\rm DN}_{G} ]
\nonumber \\ & \qquad\qquad + \sum_{q=u,d,s,c,b_2} [g^{(1)\rm DN}_q+g^{(2)\rm DN}_q] \, \Big( q(2,\mu^2) + \overline{q}(2,\mu^2) \Big) \Bigg] ,
\\
f^{\rm DN}_{b(2)+\overline{b}(2)} & = m_N \left[ \frac{2}{27} f_{TG} \sum_{Q=c,b,t} \left( -\frac{12\pi}{\alpha_s} f^{(Q)\rm DN}_G \right)  + \sum_{q=u,d,s} f^{\rm DN}_q f_{Tq} \right]
\nonumber \\
& +
\frac{3}{4} m_N \Bigg[ G(2,\mu^2) [ g^{(1,t)\rm DN}_{G}+g^{(2,t)\rm DN}_{G} ] \nonumber \\ & \qquad\qquad + \sum_{q=u,d,s,c,b} [g^{(1)\rm DN}_q+g^{(2)\rm DN}_q] \, \Big( q(2,\mu^2) + \overline{q}(2,\mu^2) \Big) \Bigg]
,
\\
f^{\rm DN} & = \minabs \left( f^{\rm DN}_{G(2)} , \, f^{\rm DN}_{b(2)+\overline{b}(2)} \right) .
\label{eq:dnfb3}
\end{align}
Here we have defined the function
\begin{align}
\minabs(x,y) = \begin{cases} x, & \text{if $|x|\le |y|$,} \\ y, & \text{if $|y|\le |x|$.} \end{cases}
\end{align}

For the lightest sbottom case of interest to us, a separation in $a_b^2-b_b^2$ and $a_b^2+b_b^2$ allows the reader to have a clear comparison with our Eq.~(\ref{eq:fb}),
\begin{align}
\left. f \right|_{b,\,G(2)}^{\rm DN} = m_N & \Bigg[  \frac{ a_b^2-b_b^2 }{4} m_b \Bigg( \frac{2}{27} \, f_{TG}\, f_{D}^{(b)} + \frac{3}{4} \,G(2,\mu^2) \, g_{D}^{(b)} \Bigg)
\nonumber \\ &
+ \frac{ a_b^2+b_b^2}{4} m_\chi \Bigg( \frac{2}{27} \, f_{TG}\, f_{S}^{(b)} +
\frac{3}{4} \,G(2,\mu^2) \, g_{S}^{(b)}  \Bigg) \Bigg] ,
\label{eq:dnfb_gluon_twist}
\end{align}
\begin{align}
\left. f \right|_{b,\,[b(2)+\overline{b}(2)]}^{\rm DN}  = m_N & \Bigg[  \frac{ a_b^2-b_b^2 }{4} m_b \Bigg( \frac{2}{27} \, f_{TG}\, f_{D}^{(b)} \Bigg)
\nonumber \\ &
+ \frac{ a_b^2+b_b^2}{4} m_\chi \Bigg( \frac{2}{27} \, f_{TG}\, f_{S}^{(b)} +
 \frac{3}{2} \frac{b(2,\mu^2) + \overline{b}(2,\mu^2)}{[m_{\tilde{q}}^2 - (m_\chi+m_q)^2]^2} \Bigg)\Bigg] ,
\label{eq:dnfb_quark_twist}
\\
\left. f \right|_{b}^{\rm DN} & = \minabs \left( \left. f \right|_{b,\,G(2)}^{\rm DN} , \, \left. f \right|_{b,\,[b(2)+\overline{b}(2)]}^{\rm DN} \right) .
\label{eq:dnfb2}
\end{align}

\section{Hisano, Ishiwata and Nagata}

HIN's effective Lagrangian is almost the same as ours, but it contains the heavy quark $c,b,t$ terms in $m_Q \overline{Q} Q$, which we have replaced with the $G^{a\mu\nu} G^{a}_{\mu\nu}$ operator using Eq.~(\ref{eq:svz}). More precisely, HIN's Lagrangian in their equation (1) does not contain the heavy quark twist-2 operator $\mathcal{O}^{(2)}_{Q\mu\nu} $, but they re-introduce it for $c$ and $b$ quarks in their equation (6) and footnote 1.

Since we have adopted the same notation as HIN for the coefficients in the effective neutralino--parton Lagrangian, we can read them directly from their paper:
\begin{align}
f_q^{\rm HIN} & = - \frac{1}{4m_q} \frac{\aq ^2-\bq^2}{m_{\tilde{q}}^2 - m_\chi^2} + \frac{m_\chi}{8} \frac{\aq ^2+\bq^2}{(m_{\tilde{q}}^2 - m_\chi^2)^2} ,
\label{eq:his_fq}
\\
g^{(1)\rm HIN}_q + g^{(2)\rm HIN}_q & = \frac{m_\chi}{2} \frac{\aq ^2+\bq^2}{(m_{\tilde{q}}^2 - m_\chi^2)^2}
\label{eq:his_gq}
\\
-\frac{12\pi}{\alpha_s} f^{(Q)\rm HIN}_G & = -3 \left[  \frac{\aQ^2+\bQ^2}{4} m_\chi (f^s_{+} + f^l_{+} ) + \frac{\aQ^2-\bQ^2}{4} m_Q (f^s_{-} + f^l_{-} )  \right],
\label{eq:hinfG}
\\
g^{(1,Q)\rm HIN}_G + g^{(2,Q)\rm HIN}_G & \simeq 0 .
\end{align}
Regarding these expressions, HIN take the zero quark mass limit in $f_q$ and $g^{(1)}_q + g^{(2)}_q $, and neglect the gluon twist-2 term $g^{(1,Q)\rm HIN}_G + g^{(2,Q)\rm HIN}_G$ because they are suppressed by $\alpha_s$ with respect to the other terms (see Eqs.~(\ref{eq:fpmgpm3}) and~(\ref{eq:fpmgpm4})). The functions $f_{\pm}^{l,s}$ are defined by HIN as
\begin{align}
f_{+}^s(m_{\tilde{Q}},m_Q,m_\chi) & = m_{\tilde{Q}}^2 \left( B_0^{(1,4)} + B_1^{(1,4)} \right) ,
\label{eq:hinfp}
\\
f_{+}^l(m_{\tilde{Q}},m_Q,m_\chi) & = m_{Q}^2 \left( B_0^{(4,1)} + B_1^{(4,1)} \right) ,
\\
f_{-}^s(m_{\tilde{Q}},m_Q,m_\chi) & = m_{\tilde{Q}}^2 B_0^{(1,4)} ,
\\
f_{-}^l(m_{\tilde{Q}},m_Q,m_\chi) & = B_0^{(3,1)} + m_{Q}^2 B_0^{(4,1)} .
\label{eq:hinfm}
\end{align}
The expressions of the loop integrals $ B_i^{(n,m)}(m_{\tilde{Q}},m_Q,m_\chi) $ defined by HIN are provided for completeness in the Appendix, where we also give their analytic continuation for $m_\chi < m_{\tilde{Q}}$.

Using the analytic expressions of the DN loop integrals $I_n$ and of the HIN loop integrals $B_i^{(n,m)}$, we find the relations
\begin{align}
f^s_{+}+f^l_{+} & = \frac{1}{2} I_2 - \frac{1}{3} m_\chi^2 I_4 - \frac{1}{6} I_5 , \label{eq:drees_nojiri_split1}
\\
f^s_{-}+f^l_{-} & = \frac{1}{2} I_1 - \frac{1}{3} m_\chi^2 I_3.
\label{eq:drees_nojiri_split2}
\end{align}
As a consequence, we have established that the DN and HIN expressions for $f_G^{(Q)}$ in Eqs.~(\ref{eq:dnfG}) and (\ref{eq:hinfG})  are identical,
\begin{align}
f_G^{(Q)\rm HIN} = f_G^{(Q)\rm DN} .
\end{align}

A key point of HIN's paper is the separation of the coefficients into short-- and long--distance parts $f_{\pm}^s$ and $f_{\pm}^l$, respectively. This separation arises from a classification of the loop integrals $B_i^{(n,m)}$ into short--distance integrals $B_i^{(1,4)}$ and long--distance integrals $B_i^{(4,1)}$ and $B_i^{(3,1)}$, distinguished based on their behavior as $m_q\to0$. In the Fock-Schwinger gauge used by HIN for the background gluon field, the long-- and short--distance integrals arise from different loop diagrams, but they cannot be easily separated
using the gauge and loop integrals in DN (see Eq.~(\ref{eq:drees_nojiri_split1})-(\ref{eq:drees_nojiri_split2})). The long--distance integrals are dominated by the mass scale of the
external quark, and arise from the diagram in Fig.~\ref{fig:dia2}(a). The
short--distance integrals are dominated by the mass scale of a heavy
particle, such as the WIMP or the squark, and arise from the diagram in
Fig.~\ref{fig:dia2}(c). The other two diagrams in Fig.~\ref{fig:dia2} vanish in the Fock-Schwinger gauge.  The long--distance loop integrals of
Fig.~\ref{fig:dia2}(a) contain more quark propagators and fewer
squark propagators compared to the short--distance loop integrals of Fig.~\ref{fig:dia2}(c),
so the former diverge faster than the latter when
$m_q\rightarrow 0$, and vanish faster when $m_{\tilde{q}}\rightarrow
\infty$. For this reason, it is the long--distance integrals that dominate the heavy squark limit.
 In any case, for a heavy quark like the bottom quark we focus on, HIN argue that the long-- and short--distance contributions must be added together, as in Eq.~(\ref{eq:hinfG}).

To summarize, HIN's expression for $f$ for a heavy quark is
\begin{align}
\left. f \right|_Q^{\rm HIN} = & \, m_N \left[ \frac{2}{27} f_{TG} \left( - \frac{12\pi}{\alpha_s} f^{(Q)\rm HIN}_G \right)  +  \frac{3}{4} \Big( g^{(1)\rm HIN}_{Q}+g^{(2)\rm HIN}_{Q} \Big)\, \Big[ Q(2,\mu^2) + \overline{Q}(2,\mu^2) \Big] \right].
\end{align}
Notice that HIN integrated out the heavy quark in the trace term but not in the twist-2 term. This is an example of the hybrid approach we have argued against above.

Separating the $\aQ^2-\bQ^2$ and $\aQ^2+\bQ^2$ terms, and specializing to our case of the lightest sbottom, facilitates the comparison with our Eq.~(\ref{eq:fb}) and DN's expression in Eq.~(\ref{eq:dnfb2}),
\begin{align}
\left. f \right|_b^{\rm HIN} = m_N & \left[  \frac{ a_b^2-b_b^2 }{4} m_b \left( \frac{2}{27} \, f_{TG}\, f_{D}^{(b)}  \right)
+ \frac{ a_b^2+b_b^2}{4} m_\chi \Bigg( \frac{2}{27} \, f_{TG}\, f_{S}^{(b)} + \frac{3}{2} \frac{b(2,\mu^2) + \overline{b}(2,\mu^2)}{[m_{\tilde{b}}^2 - m_\chi^2]^2} \Bigg) \right] .
\label{eq:hinfb}
\end{align}
Notice that the last term diverges at $m_{\tilde{b}}=m_\chi$.

\section{\Mo}
\label{sec:micromega}
To obtain the scattering amplitude and scattering cross section in \mo\ (version 3.1), we have defined a new model that extends the Standard Model by the addition of a Majorana particle of spin 1/2 (the ``neutralino'') and a scalar particle of spin 0 (the ``sbottom'') coupled to the bottom quark through the Lagrangian in Eq.~(\ref{eq:lagrangian}). For this purpose, we have written a \mo\ particle file {\tt work/models/prtcls1.mdl} defining a bottom squark and a neutralino, and a \mo\ model file {\tt work/models/lgrng1.mdl} containing squark--gluon couplings and neutralino--quark--squark couplings. We have then modified the main program provided with the \mo\ distribution in such a way that only the {\tt CDM\_NUCLEON} module remains. Our modified main program assigns values to the squark mass $m_{\tilde{q}}$ and to the coupling coefficients $\aq$ and $\bq$, and then tabulates (a) the \mo\ spin-independent scattering amplitudes {\tt pA0} and {\tt nA0}, which are equal to our function $f$ for protons and neutrons, respectively, and (b) the scattering cross sections {\tt xsp} and {\tt xsn} for protons and neutrons, for which precoded \mo\ expressions in terms of {\tt pA0} and {\tt nA0} are used.

In the default option, \mo\ uses a special numerical technique described in~\cite{micromegas_direct} to compute the effective Lagrangian coefficients $f_q$ and $g_q^{(i)}$. Notice that these neutralino--quark coefficients are used for both light and heavy quarks, i.e.\ \mo\ in the default option does no integrate out the heavy quarks  $Q=c,b,t$, while we have integrated them out and use the coefficients of the neutralino--gluon effective Lagrangian.  For the specific case of the fundamental Lagrangian $\mathcal{L}_{\tilde{q}q\chi} $ in Eq.~(\ref{eq:lagrangian}), Ref.~\cite{micromegas_direct} quotes a neutralino-quark scattering amplitude at zero-momentum-transfer equal to
\begin{align}
A = \frac{1}{4} \left[ \frac{\bq^2}{m_{\tilde{q}}^2-(m_\chi+m_q)^2} - \frac{\aq^2}{m_{\tilde{q}}^2-(m_\chi-m_q)^2} \right].
\label{eq:Atreelevel}
\end{align}
From the information in Ref.~\cite{micromegas_direct} we therefore deduce the following \mo\ coefficients in the default option (here $q=u,d,s,c,b,t$)
\begin{align}
f_q^{\rm MO,default} & = \frac{1}{4m_q} \left[ \frac{\bq^2}{m_{\tilde{q}}^2-(m_\chi+m_q)^2} - \frac{\aq^2}{m_{\tilde{q}}^2-(m_\chi-m_q)^2} \right],
\label{eq:MO1}
\\
g^{(1)\rm MO,default}_q + g^{(2)\rm MO,default}_q & = g^{\rm MO}(m_{\tilde{q}},m_q,m_\chi)
\\
-\frac{12\pi}{\alpha_s} f^{(Q)\rm MO,default}_G & =0 ,\label{eq:hinfG}
\\
g^{(1,Q)\rm MO,default}_G + g^{(2,Q)\rm MO,default}_G & = 0.
\label{eq:MO4}
\end{align}
Here we have indicated that the twist-2 quark coefficients are a function $g^{\rm MO}$ of the squark, quark, and neutralino masses, but we have been unable to compute the analytic form of this function from the explanations in Ref.~\cite{micromegas_direct}.

\Mo\  also provides an option {\tt FeScLoop} to replace the tree-level amplitude in Eq.~(\ref{eq:Atreelevel}) with the Drees-Nojiri gluon trace terms. Ref.~\cite{micromegas_direct} advocates the use of this option in the case $m_{\tilde{q}} < m_\chi + m_q$. With the  {\tt FeScLoop} option, \mo\ omits the default squark coefficients in Eqs.~(\ref{eq:MO1})-(\ref{eq:MO4}), and replaces them with following expressions,
 \begin{align}
f^{\rm MO,FeScLoop}_{q} & = \frac{\aq^2-\bq^2}{4} \left( m_\chi^2 I_3 - \frac{3}{2} m_q I_1 \right) \nonumber \\ & + \frac{\aq^2+\bq^2}{4} m_\chi \left( m_\chi^2 I_4+ \frac{1}{2} I_5 -\frac{3}{2} I_2 \right) ,
\label{eq:fescloop1}
\\
g^{(1)\rm MO,FeScLoop}_{q} + g^{(2)\rm MO,FeScLoop}_{q} & = 0,
\\[1ex]
-\frac{12\pi}{\alpha_s} f^{(Q)\rm MO,FeScLoop}_{G} & = 0,
\\
g^{(1,Q)\rm MO,FeScLoop}_{G} + g^{(2,Q)\rm MO,FeScLoop}_{G} & = 0.
\label{eq:fescloop4}
\end{align}
These expressions, which are given in Eqs.~(A-3)-(A-5) in Ref.~\cite{micromegas_direct} and are used in the \mo\ code version 3.1 (and earlier versions), have an incorrect $\aq^2-\bq^2$ coefficient, in the sense that the factor $m_\chi^2 I_3 - \frac{3}{2} m_q I_1$ should have been $m_q ( m_\chi^2 I_3 - \frac{3}{2} I_1 ) $.

Finally, \mo\ provides a function {\tt MSSMDDtest} that offers the Drees-Nojiri formulas from either the DN neutralino--quark or the DN neutralino--gluon effective Lagrangian, with the addition of QCD and SUSY-QCD corrections. However, this routine requires setting up the complete MSSM model, and we were unable to choose the MSSM parameters to match our calculations with the Lagrangian in Eq.~(\ref{eq:lagrangian}). We have therefore omitted a comparison with this option.

We end this section by writing the \mo\ formula for the bottom quark
$f|_b$ in the default option, separated in $a_b^2-b_b^2$ and
$a_b^2+b_b^2$, so that the reader can easily compare it with those in
the other sections (here $g^{\rm MO}_{D}$ and $g^{\rm MO}_{S}$ denote
the $a_b^2-b_b^2$ and $a_b^2+b_b^2$ parts of the function $g^{\rm MO}$
introduced above).
\begin{align}
\left. f \right|_b^{\rm MO,default} = m_N & \Bigg\{  \frac{ a_b^2-b_b^2 }{4} m_b \Bigg( \frac{2}{27} \, f_{TG}\, \frac{1}{2m_b^2} \left[ - \frac{1}{m_{\tilde{b}}^2-(m_\chi+m_b)^2} - \frac{1}{m_{\tilde{b}}^2-(m_\chi-m_b)^2} \right]
\nonumber \\ & \qquad \qquad \qquad + \frac{3}{m_\chi} \Big[ b(2,\mu^2) + \overline{b}(2,\mu^2) \big] g^{\rm MO}_{D} \Bigg)
\nonumber \\ &
+ \frac{ a_b^2+b_b^2}{4} m_\chi \Bigg( \frac{2}{27} \, f_{TG}\,\frac{1}{2m_bm_\chi} \left[  \frac{1}{m_{\tilde{b}}^2-(m_\chi+m_b)^2} - \frac{1}{m_{\tilde{b}}^2-(m_\chi-m_b)^2} \right]
\nonumber \\ & \qquad \qquad \qquad + \frac{3}{m_\chi} \Big[ b(2,\mu^2) + \overline{b}(2,\mu^2) \big] g^{\rm MO}_{S} \Bigg) \Bigg\} .
\label{eq:hinfb}
\end{align}
Notice that $\left. f \right|_b^{\rm MO,default} $ diverges when $m_{\tilde{b}} = m_\chi + m_b$, as well as when $m_{\tilde{b}} = m_\chi - m_b$. However, \mo\ does not allow the user to access the region $m_\chi < m_{\tilde{b}}$.

\section{DarkSUSY}
\label{sec:darksusy}

DarkSUSY (version 5.1) allows several options for the calculation of the scattering cross section, letting the user specify whether to include the squark poles or not and whether to use the Drees-Nojiri expressions or the limiting heavy-squark expressions.
While a standard use of DarkSUSY would give the scattering cross section, or even the higher-level scattering rate, here we want to extract the scattering amplitudes.

In DarkSUSY, there are four scattering amplitudes available, one each for the four combinations of spin-dependent and spin-independent scattering off protons and neutrons. They are called {\tt gps, gns, gpa, gna} for spin-independent off proton and neutron and spin-dependent off proton and neutron, respectively. They are computed by the function {\tt dsddgpgn}. The latter function requires values for the MSSM particle masses and coupling constants. Since we want to obtain amplitudes for simple values of $\aq$ and $\bq$, we write the DarkSUSY squark-neutralino-quark couplings $g_{R\tilde{q}\chi q}$ and $g_{L\tilde{q}\chi q}$ in terms of $\aq$ and $\bq$. By definition,
\begin{align}
\mathcal{L}_{\tilde{q}\chi q} =  \tilde{q}^* \, \overline{\chi} (g_{L\tilde{q}\chi q} P_L + g_{R\tilde{q}\chi q} P_R) q + \text{h.c.} ,
\end{align}
where $P_L = (1-\gamma_5)/2$ and $P_R=(1+\gamma_5)/2$. Comparing with Eq.~(\ref{eq:lagrangian})  gives the relations
\begin{align}
g_{L\tilde{q}\chi q}  = a^*_{\tilde{q}} + b^*_{\tilde{q}}, \\
g_{R\tilde{q}\chi q}  = a^*_{\tilde{q}} - b^*_{\tilde{q}}.
\end{align}

With the default option, DarkSUSY uses the heavy-squark limit expressions for both light and heavy quarks,
\begin{align}
f^{\rm DS,default}_q  & = - \frac{\aq ^2-\bq^2}{4m_qm_{\tilde{q}}^2}  , & (q=u,d,s,c,b,t)
\label{eq:DSdefault1}
\\
-\frac{12\pi}{\alpha_s} f_G^{(Q)\rm DS,default} & = g^{(i,Q)}_G = g^{(i)\rm DS,default}_q = 0  , &
\label{eq:DSdefault2}
\end{align}
With the {\tt pole} option, DarkSUSY introduces a pole into the propagators of the heavy-squark limit expressions,
\begin{align}
f^{\rm DS,pole}_q  & = - \frac{\aq ^2-\bq^2}{4m_q[m_{\tilde{q}}^2-(m_\chi+m_q)^2]}  , & (q=u,d,s,c,b,t)
\label{eq:DSpole1}
\\
-\frac{12\pi}{\alpha_s} f_G^{(Q)\rm DS,pole} & = g^{(i)\rm DS,pole}_q = g^{(i)\rm DS,pole}_G = 0 .
\label{eq:DSpole2}
\end{align}
Finally, with the {\tt dn1} option, DarkSUSY uses the DN formulas in Section~\ref{sec:DN}, however it has its own prescription for the bottom quark. Instead of DN's prescription described at the end of Section~\ref{sec:DN}, Eqs.~(\ref{eq:dnfb1})-(\ref{eq:dnfb2}), where the smallest (in absolute value) total amplitude is selected, DarkSUSY selects the smallest (in absolute value) of the twist-2 amplitudes. Collecting all formulas, the DarkSUSY option {\tt dn1} uses
\begin{align}
f^{\rm DS,dn1} & = m_N \left[ \frac{2}{27} f_{TG} \sum_{Q=c,b,t} \left( -\frac{12\pi}{\alpha_s} f^{(Q)\rm DN}_G \right)  + \sum_{q=u,d,s} f^{\rm DN}_q f_{Tq} \right]
\nonumber \\
& +
\frac{3}{4} m_N \Bigg[ G(2,\mu^2) [ g^{(1,t)\rm DN}_{G}+g^{(2,t)\rm DN}_{G} ] \,\,\, + \sum_{q=u,d,s,c} [g^{(1)\rm DN}_q+g^{(2)\rm DN}_q] \, \Big( q(2,\mu^2) + \overline{q}(2,\mu^2) \Big)
\nonumber \\
& \qquad \qquad +
\minabs\!\Big[ G(2,\mu^2) [ g^{(1,b)\rm DN}_{G}+g^{(2,b)\rm DN}_{G} ] , \, [g^{(1)\rm DN}_b+g^{(2)\rm DN}_b] \, \Big( b(2,\mu^2) + \overline{b}(2,\mu^2) \Big] \Bigg] .
\label{eq:dsf}
\end{align}
In particular, for the bottom quark of interest to us,
\begin{align}
\left. f \right|_b^{\rm DS,default} & = - m_N \frac{2}{27} f_{TG} \frac{a_b^2-b_b^2}{4m_bm_{\tilde{b}}^2} ,
\\
\left. f \right|_b^{\rm DS,pole} & = - m_N \frac{2}{27} f_{TG} \frac{a_b^2-b_b^2}{4m_b[m_{\tilde{b}}^2-(m_\chi+m_b)^2]} ,
\end{align}
and
\begin{align}
\left. f \right|_b^{\rm DS,dn1} = m_N \Bigg\{ & \frac{ a_b^2-b_b^2 }{4} m_b \left[ \frac{2}{27} \, f_{TG}\, f_{D}^{(b)} + \frac{3}{4} \minabs\!\Big( G(2,\mu^2) \, g_{D}^{(b)}, 0 \Big) \right]
\nonumber \\ &
+ \frac{ a_b^2+b_b^2}{4} m_\chi \Bigg[ \frac{2}{27} \, f_{TG}\, f_{S}^{(b)}
+ \frac{3}{4} \minabs\!\bigg[ G(2,\mu^2) \, g_{S}^{(b)},
\nonumber\\ & \qquad\qquad\qquad \frac{m_\chi}{2} \frac{\aq ^2+\bq^2}{[m_{\tilde{b}}^2 - (m_\chi+m_b)^2]^2} \Big( b(2,\mu^2)+\overline{b}(2,\mu^2) \Big)  \bigg] \Bigg] \Bigg\} .
\end{align}

\section{Quantitative analysis}
\label{sec:results}
%%%%%%%%%%%%%%%%%%%%%%%%%%%%%%%%%%%%%%%%%%%%%%%%%%%%%%%%%%%%%%%
\begin{figure}
\begin{center}
\includegraphics[width=0.80\linewidth]{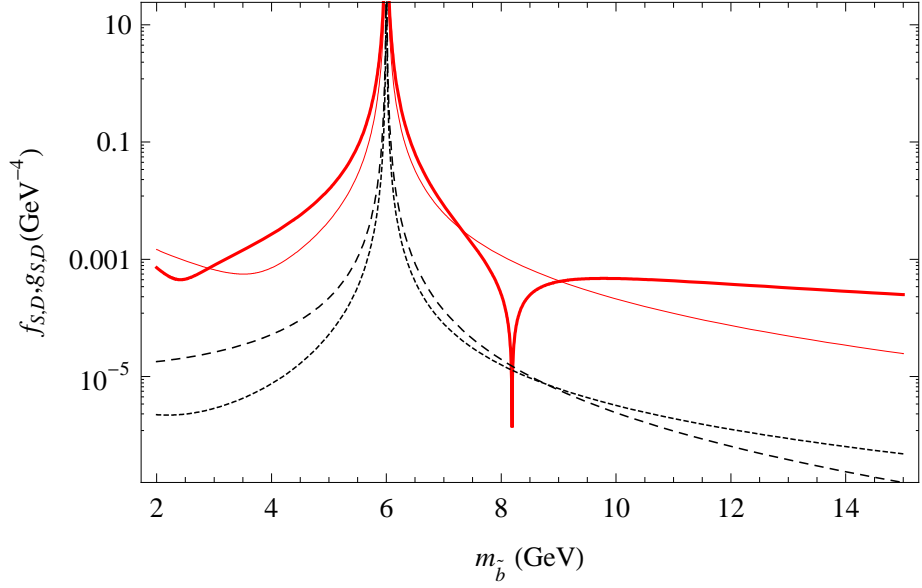}
\end{center}
\caption{The amplitudes $f_{S}^{(Q)}$, $f_{D}^{(Q)}$, $g_{S}^{(Q)}$, $g_{D}^{(Q)}$ defined in
  Eqs.(\protect\ref{eq:fpmgpm1}--\protect\ref{eq:fpmgpm4}) are plotted
  (in absolute value) as a function of $m_{\tilde{b}}$ for
  $m_{\chi}=10$ GeV and $m_b$=4 GeV.  Thick red solid line: $f_{D}^{(Q)}$;
  thin red solid line: $f_{S}^{(Q)}$; black long--dashed line: $g_{D}^{(Q)}$; black short--dashed line: $g_{S}^{(Q)}$.  All the amplitudes show a single pole at
  $m_{\tilde{b}}=m_{\chi}-m_b$, while they are finite everywhere else,
  in particular at $m_{\tilde{b}}=m_{\chi}$ and at
  $m_{\tilde{b}}=m_{\chi}+m_b$. Notice that the functions $g_{S}^{(Q)}$ and
  $g_{D}^{(Q)}$ (black dashed lines) are ${\cal O}(\alpha_s)$ suppressed compared to $f_{S}^{(Q)}$ and
  $f_{D}^{(Q)}$ (red solid lines), as seen in Eqs.~(\ref{eq:fpmgpm3},\ref{eq:fpmgpm4}).
\label{fig:amplitudes}}
\end{figure}
%%%%%%%%%%%%%%%%%%%%%%%%%%%%%%%%%%%%%%%%%%%%%%%%%%%%%%%%%%%%%%%%

%%%%%%%%%%%%%%%%%%%%%%%%%%%%%%%%%%%%%%%%%%%%%%%%%%%%%%%
\begin{figure}[t]
\begin{center}
\includegraphics[width=0.80\columnwidth]{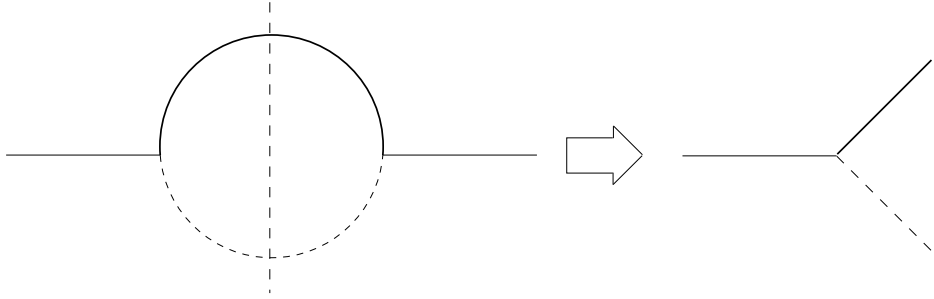}
\end{center}
\caption{In the limit of zero gluon momenta the diagrams of
  Fig. \protect\ref{fig:dia2} have the same analytic behavior as the
  neutralino self energy, with a single cut in correspondence of the
  opening of the decay process $\chi \rightarrow \tilde{q}+q$, when
  $\mc=\msq+m_q$. As in Fig. \protect\ref{fig:dia2} neutralinos are
  shown with thin solid lines, quarks with thick solid lines and
  squarks with dashed lines.}
\label{fig:cutkoski}
\end{figure}
%%%%%%%%%%%%%%%%%%%%%%%%%%%%%%%%%%%%%%%%%%%%%%%%%%%%%%%

%%%%%%%%%%%%%%%%%%%%%%%%%%%%%%%%%%%%%%%%%%%%%%%%%%%%%%%%%%%%%%%
\begin{figure}[h]
\begin{center}
\includegraphics[width=0.80\linewidth]{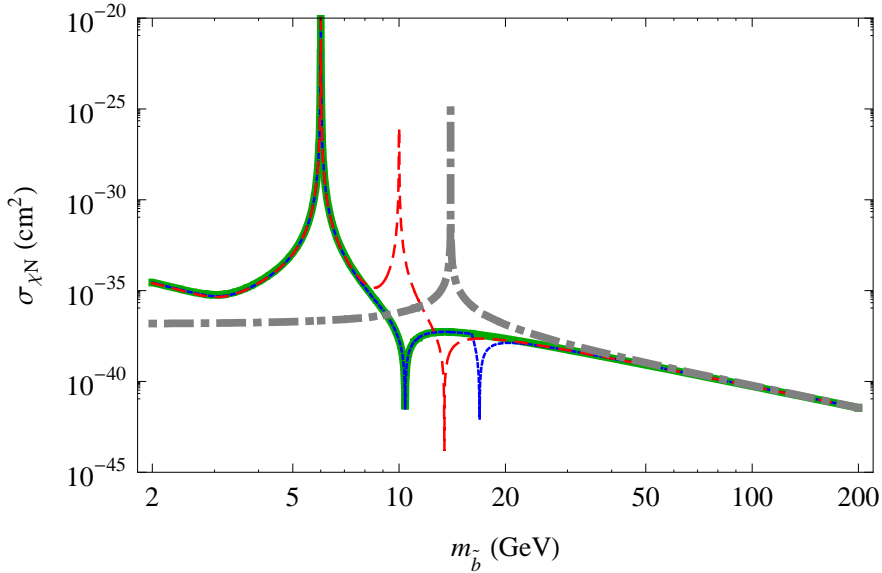}
\end{center}
\caption{Neutralino--nucleon cross section as a function of the bottom squark
  mass calculated assuming $\aq=1$, $\bq=0$ and $\tilde{q}=\tilde{b}$.
  Thick green solid line: cross section calculated using the
  transition amplitude given in Eq.~(\protect\ref{eq:fb}); thin blue
  dotted line: the same using the transition amplitude of
  Eq.~(\protect\ref{eq:dnfb2}); red dashed line: the same using the
  transition amplitude of Eq.~(\protect\ref{eq:hinfb}); gray
  dot-dashed line: the same using the transition amplitude given by
  the heavy--squark limit expression of
  Eq.~(\protect\ref{eq:heavy_squark}) extrapolated to lower masses
  with the ad--hoc substitution $m_{\tilde{b}}^2\rightarrow
  m_{\tilde{b}}^2-(\mc+m_b)^2$ in the propagator.}
\label{fig:sigma_direct_a1_b0}
\end{figure}
%%%%%%%%%%%%%%%%%%%%%%%%%%%%%%%%%%%%%%%%%%%%%%%%%%%%%%%%%%%%%%%%

%%%%%%%%%%%%%%%%%%%%%%%%%%%%%%%%%%%%%%%%%%%%%%%%%%%%%%%%%%%%%%%
\begin{figure}[h]
\begin{center}
\includegraphics[width=0.80\linewidth]{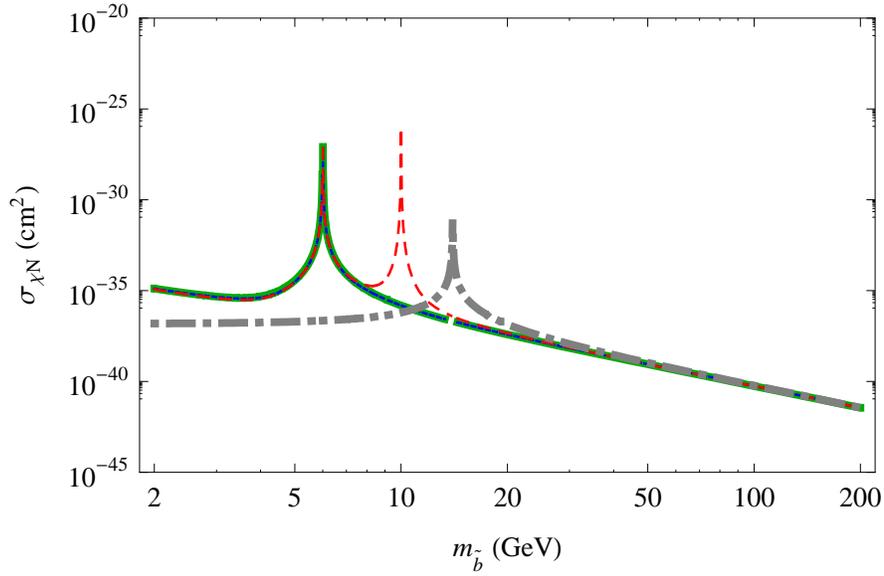}
\end{center}
\caption{Same as Fig.~\protect\ref{fig:sigma_direct_a1_b0} but for $a_{\tilde{b}}=0$, $b_{\tilde{b}}=1$.}
\label{fig:sigma_direct_a0_b1}
\end{figure}
%%%%%%%%%%%%%%%%%%%%%%%%%%%%%%%%%%%%%%%%%%%%%%%%%%%%%%%%%%%%%%%%

%%%%%%%%%%%%%%%%%%%%%%%%%%%%%%%%%%%%%%%%%%%%%%%%%%%%%%%%%%%%%%%
\begin{figure}[h]
\begin{center}
\includegraphics[width=0.80\linewidth]{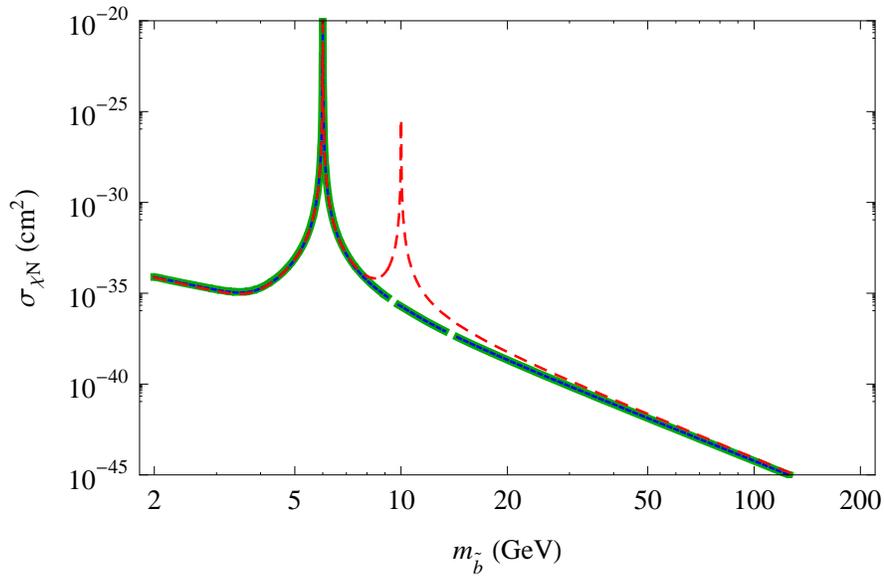}
\end{center}
\caption{Same as Fig.~\protect\ref{fig:sigma_direct_a1_b0} but for
  $a_{\tilde{b}}=1$, $b_{\tilde{b}}=1$. The curve corresponding to the
  heavy--squark approximation with modified propagator (gray dot-dashed line in
  Figs.~\protect\ref{fig:sigma_direct_a1_b0} and~\protect\ref{fig:sigma_direct_a0_b1})
  is missing, because the corresponding amplitude contains only a
  contribution proportional to $a_{\tilde{b}}^2-b_{\tilde{b}}^2$, which vanishes in this case.}
\label{fig:sigma_direct_a1_b1}
\end{figure}
%%%%%%%%%%%%%%%%%%%%%%%%%%%%%%%%%%%%%%%%%%%%%%%%%%%%%%%%%%%%%%%%

%%%%%%%%%%%%%%%%%%%%%%%%%%%%%%%%%%%%%%%%%%%%%%%%%%%%%%%%%%%%%%%%
\begin{figure}[h]
\begin{center}
\includegraphics[width=0.80\linewidth]{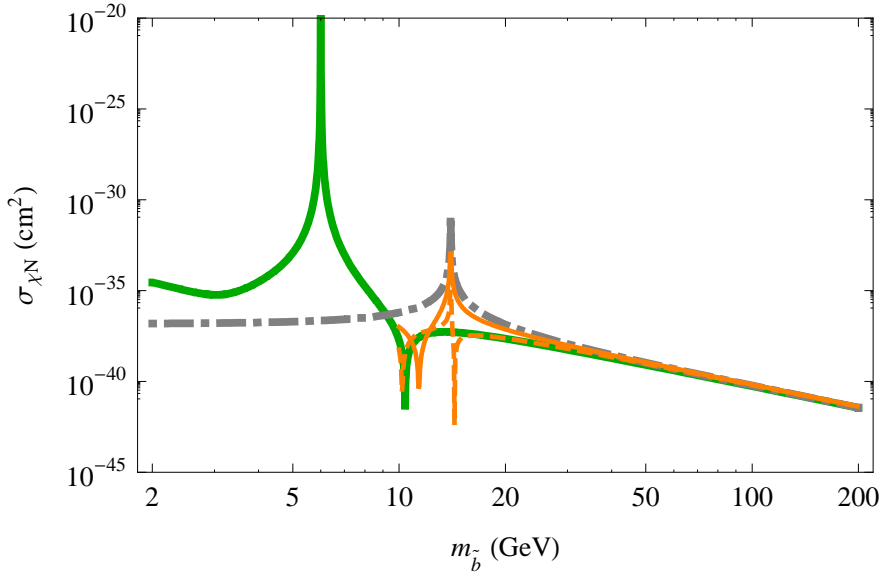}
\end{center}
\caption{ Neutralino--nucleon cross section as a function of the
  bottom squark mass calculated with the public code
  \mo~\protect\cite{micromegas} 3.1 with $\aq=1$, $\bq=0$ and
  $\tilde{q}=\tilde{b}$.  Orange thin solid line: ``default'' output
  of \mo; orange dashed line: \mo~ with the option ``FeScLoop.'' As a
  reference, the following two curves from
  Fig.~\protect\ref{fig:sigma_direct_a1_b0} are also shown. Thick
  green solid line: cross section calculated using the transition
  amplitude as given in Eq.~(\protect\ref{eq:fb}); gray dot-dashed line:
  cross section calculated using the transition amplitude given by the
  heavy--squark limit expression of Eq.~(\protect\ref{eq:heavy_squark})
  extrapolated to lower masses with the ad--hoc substitution
  $m_{\tilde{b}}^2\rightarrow m_{\tilde{b}}^2-(\mc+m_b)^2$ in the
  propagator.}
\label{fig:sigma_a1_b0_ms3}
\end{figure}
%%%%%%%%%%%%%%%%%%%%%%%%%%%%%%%%%%%%%%%%%%%%%%%%%%%%%%%%%%%%%%%%

%%%%%%%%%%%%%%%%%%%%%%%%%%%%%%%%%%%%%%%%%%%%%%%%%%%%%%%%%%%%%%%%
\begin{figure}[h]
\begin{center}
\includegraphics[width=0.80\linewidth]{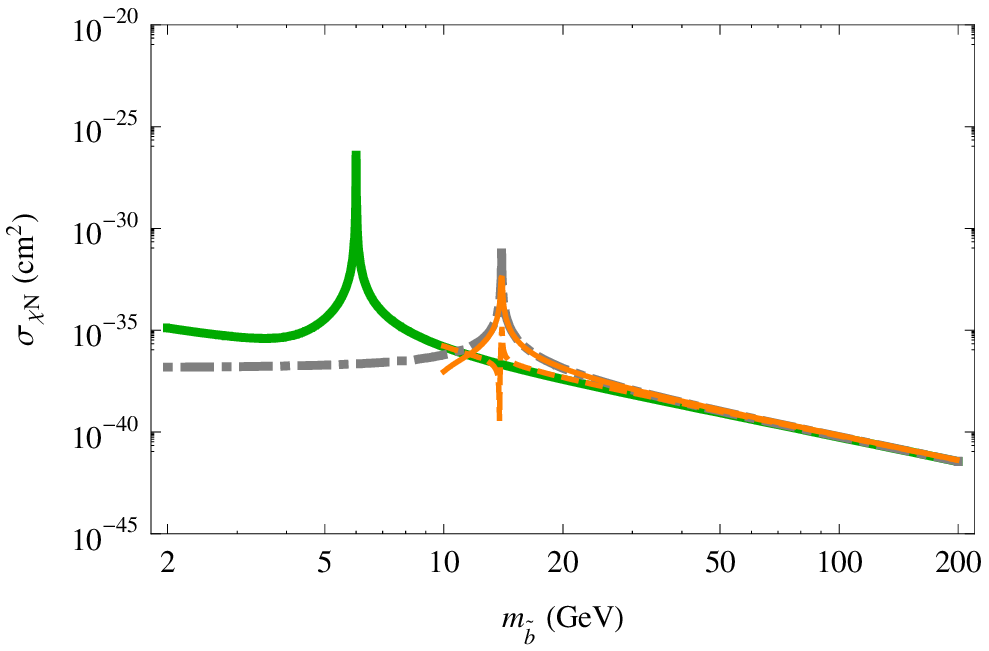}
\end{center}
\caption{Same as Fig.~\protect\ref{fig:sigma_a1_b0_ms3} but
for $\aq=1$ and $\bq=0$.}
\label{fig:sigma_a0_b1_ms3}
\end{figure}
%%%%%%%%%%%%%%%%%%%%%%%%%%%%%%%%%%%%%%%%%%%%%%%%%%%%%%%%%%%%%%%%

%%%%%%%%%%%%%%%%%%%%%%%%%%%%%%%%%%%%%%%%%%%%%%%%%%%%%%%%%%%%%%%%
\begin{figure}[h]
\begin{center}
\includegraphics[width=0.80\linewidth]{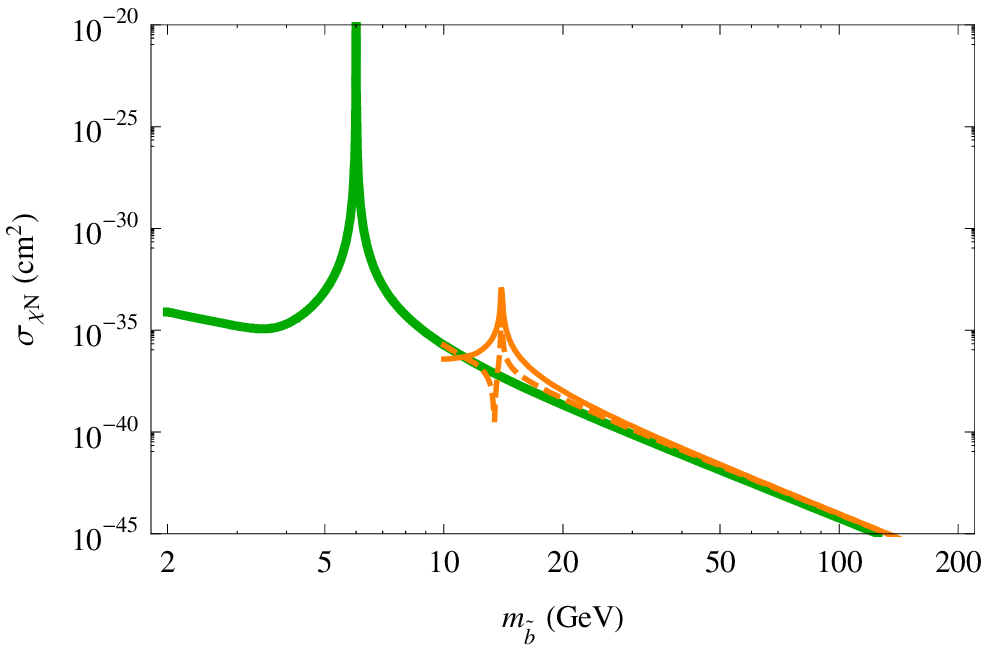}
\end{center}
\caption{Same as Fig.~\protect\ref{fig:sigma_a1_b0_ms3} but
for $\aq=1$ and $\bq=1$. The curve corresponding to the
  heavy--squark approximation with modified propagator (gray dot-dashed line in
  Figs.~\protect\ref{fig:sigma_a1_b0_ms3} and~\protect\ref{fig:sigma_a0_b1_ms3})
  is missing, because the corresponding amplitude contains only a
  contribution proportional to $a_{\tilde{b}}^2-b_{\tilde{b}}^2$, which vanishes in this case.}
\label{fig:sigma_a1_b1_ms3}
\end{figure}
%%%%%%%%%%%%%%%%%%%%%%%%%%%%%%%%%%%%%%%%%%%%%%%%%%%%%%%%%%%%%%%%

%%%%%%%%%%%%%%%%%%%%%%%%%%%%%%%%%%%%%%%%%%%%%%%%%%%%%%%%%%%%%%%%
\begin{figure}[h]
\begin{center}
\includegraphics[width=0.80\linewidth]{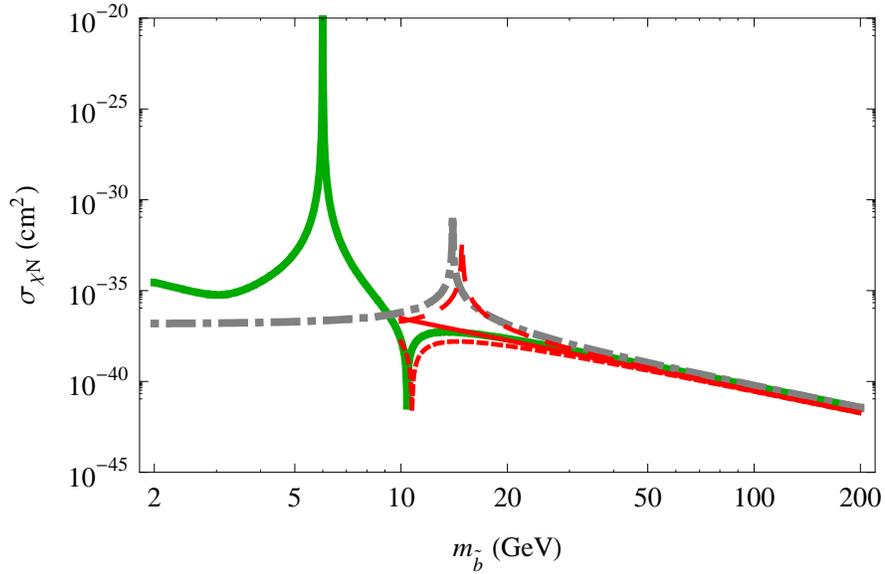}
\end{center}
\caption{ Neutralino--nucleon cross section as a function of the
  bottom squark mass calculated with the public code DarkSUSY
  5.1~\protect\cite{darksusy} with $\aq=1$, $\bq=0$ and
  $\tilde{q}=\tilde{b}$. Thin red solid line:
  DarkSUSY~\protect\cite{darksusy} with the option ``default;'' long
  red dashed line: DarkSUSY with the option ``pole;'' short red dashed
  line: DarkSUSY with the option ``Drees--Nojiri.'' The green solid
  and the gray dot-dashed lines are the same as in
  Fig. \protect\ref{fig:sigma_a1_b0_ms3}.}
\label{fig:sigma_a1_b0_ds}
\end{figure}
%%%%%%%%%%%%%%%%%%%%%%%%%%%%%%%%%%%%%%%%%%%%%%%%%%%%%%%%%%%%%%%%

%%%%%%%%%%%%%%%%%%%%%%%%%%%%%%%%%%%%%%%%%%%%%%%%%%%%%%%%%%%%%%%%
\begin{figure}[h]
\begin{center}
\includegraphics[width=0.80\linewidth]{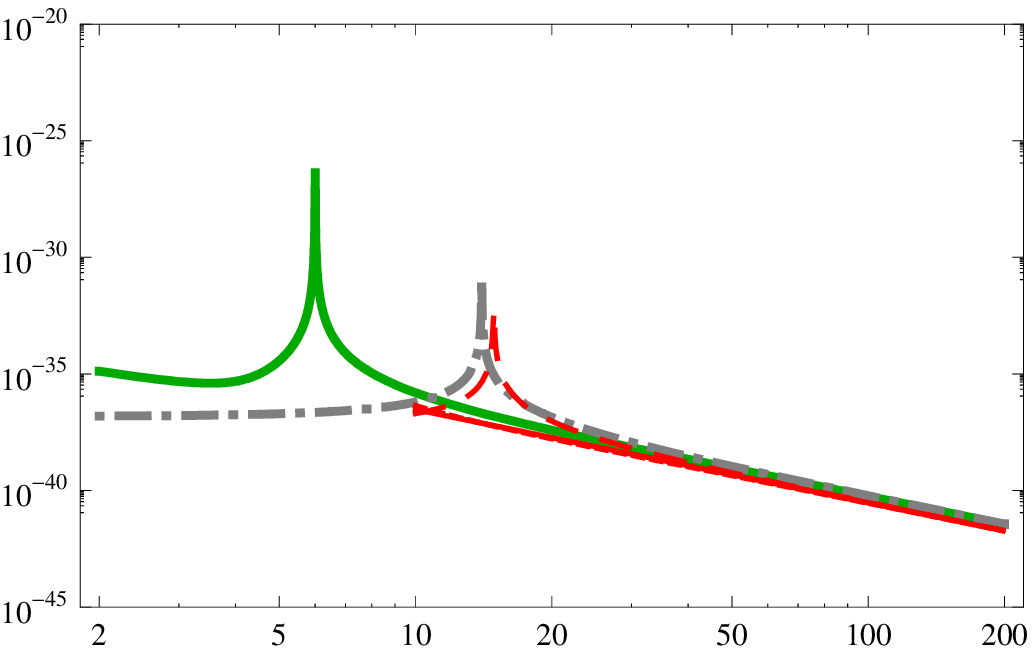}
\end{center}
\caption{Same as Fig.~\protect\ref{fig:sigma_a1_b0_ds} but
for $\aq=1$ and $\bq=0$. }
\label{fig:sigma_a0_b1_ds}
\end{figure}
%%%%%%%%%%%%%%%%%%%%%%%%%%%%%%%%%%%%%%%%%%%%%%%%%%%%%%%%%%%%%%%%

%%%%%%%%%%%%%%%%%%%%%%%%%%%%%%%%%%%%%%%%%%%%%%%%%%%%%%%%%%%%%%%%
\begin{figure}[h]
\begin{center}
\includegraphics[width=0.80\linewidth]{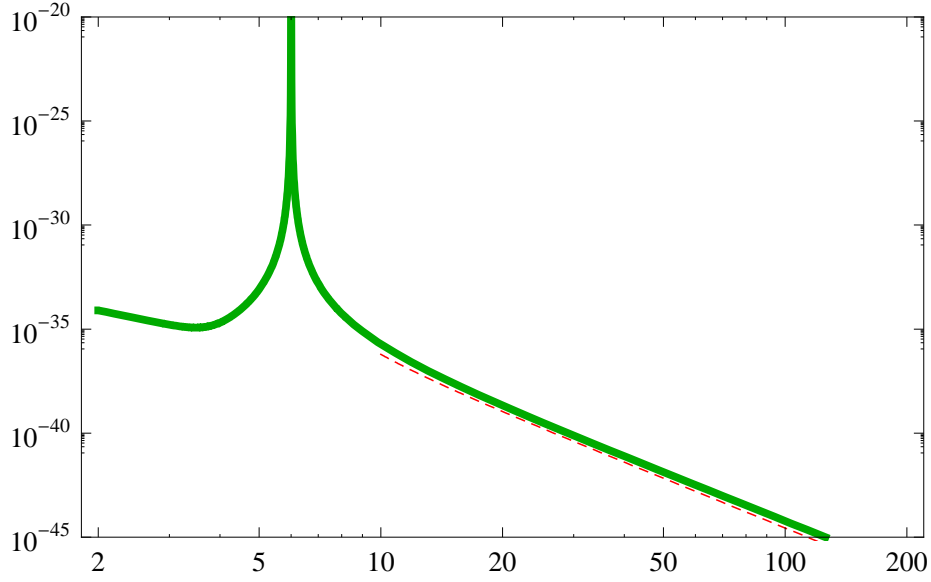}
\end{center}
\caption{Same as Fig.~\protect\ref{fig:sigma_a1_b0_ds} but for $\aq=1$
  and $\bq=1$. In this case the options ``default'' and ``pole''
  vanish, because in both cases the cross section is proportional to
  $\aq^2-\bq^2$. For the same reason, the curve corresponding to the
  heavy--squark approximation with modified propagator (gray
  dot-dashed line in Figs.~\protect\ref{fig:sigma_a1_b0_ds}
  and~\protect\ref{fig:sigma_a0_b1_ds}) is also missing.}
\label{fig:sigma_a1_b1_ds}
\end{figure}
%%%%%%%%%%%%%%%%%%%%%%%%%%%%%%%%%%%%%%%%%%%%%%%%%%%%%%%%%%%%%%%%

%%%%%%%%%%%%%%%%%%%%%%%%%%%%%%%%%%%%%%%%%%%%%%%%%%%%%%%%%%%%%%%
\begin{figure}
\begin{center}
\includegraphics[width=0.60\linewidth]{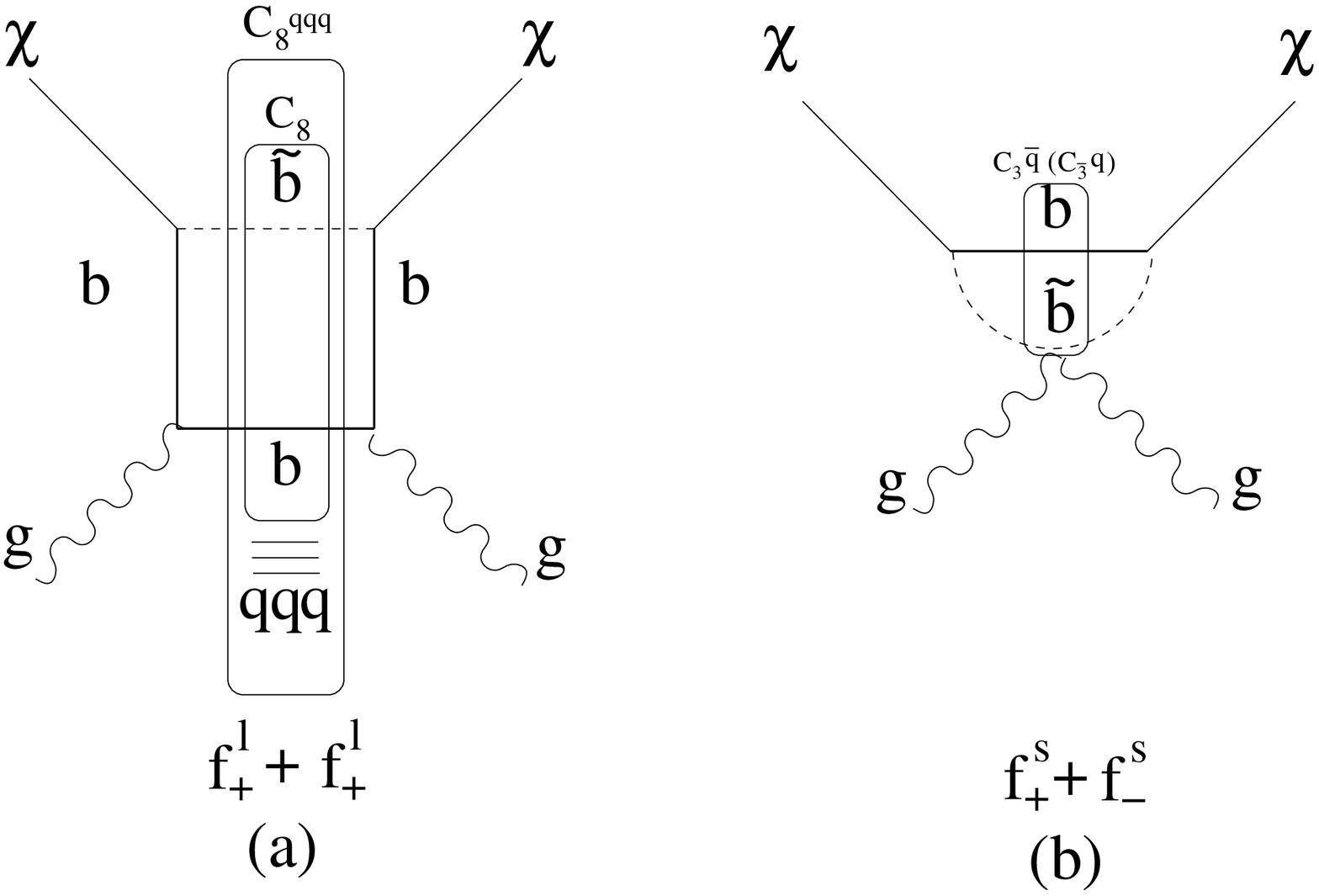}
\end{center}
\caption{ (a) Loop diagram generating the long--distance amplitudes
  $f_{+}^l+f_{-}^l$; (b) loop diagram generating the short--distance
  amplitudes $f_{+}^s+f_{-}^s$.  In both cases the pole in
  the amplitude at $m_{\tilde{q}} = m_\chi - m_q$ can be interpreted as the formation of a resonance with mass $m_R\simeq
  m_{\tilde{b}}+m_b$: a
  color--singlet R--hadron $C_8qqq$ in case (a) or a $C_3q$
  ($C_{\bar{3}}q$) state in case (b).  \label{fig:long_and_short}}
\end{figure}
%%%%%%%%%%%%%%%%%%%%%%%%%%%%%%%%%%%%%%%%%%%%%%%%%%%%%%%%%%%%%%%%

In this numerical Section we simplify the discussion by assuming that only the lightest bottom squark $\tilde{b}$ contributes to the neutralino--nucleus cross section.

As reviewed in the previous Sections, several expressions exist in the
literature for the neutralino--nucleon scattering amplitude through
squark exchange. The effective approach of tree-level
neutralino--quark scattering (Fig.~\ref{fig:dia1}), properly
calculable in the limit of heavy squarks, fails in the domain
$m_{\tilde{q}}\rightarrow m_{\chi}$, where resonances at
$m_{\tilde{q}}=m_\chi\pm m_q$ or $m_{\tilde{q}}=m_\chi$ are introduced
according to various ad hoc recipes. On the other hand, the
loop--integral approach of Fig.~\ref{fig:dia2} (neutralino--gluon
amplitudes) is reliable as long as the quark running in the loop is
heavy enough for the calculation to be perturbative, which is the case
for the bottom quark.

When the heavy quark is integrated out,  the cross section for neutralino--nucleon elastic scattering depends on the four
combinations of loop integrals $f_{D}^{(Q)}$, $f_{S}^{(Q)}$, $g_{D}^{(Q)}$, and $g_{S}^{(Q)}$, as displayed in Eq. (\ref{eq:fb}). We plot these four quantities in
Fig. \ref{fig:amplitudes} as functions of the sbottom
squark mass $m_{\tilde{b}}$ for the
case $m_{\chi}=10$ GeV, $m_Q=m_b= 4$~GeV. All four quantities  show only one pole at
$m_{\tilde{b}}=m_{\chi}-m_b$, while they are regular everywhere else,
in particular at $m_{\tilde{b}}=m_{\chi}$ and
$m_{\tilde{b}}=m_{\chi}+m_b$. In the appendix, we prove analytically that the loop integrals are regular at $m_{\tilde{b}}=m_\chi+m_b$.
In comparison, the HIN prescription in Eq.~(\ref{eq:hinfb}) introduces a pole at $m_{\tilde{b}}=m_\chi$ (obtained by neglecting the quark mass), the \mo\ prescription introduces poles at $m_{\tilde{b}}=m_\chi\pm m_b$, while the DN prescription in Eqs.~(\ref{eq:dnfb_gluon_twist})--(\ref{eq:dnfb2}) is set up to avoid the pole at $m_{\tilde{b}}=m_\chi+m_b$ through an ad hoc construction. Both the DN and HIN procedures lack a pole at $m_{\tilde{b}}=m_\chi-m_b$, which is instead present in the full loop calculation.

The existence of only one pole at $m_{\tilde{b}}=m_\chi-m_b$ can be understood in the
following way. The loop integrals $f_{D}^{(Q)}$, $f_{S}^{(Q)}$, $g_{D}^{(Q)}$, and $g_{S}^{(Q)}$ are
calculated from the loop diagrams shown in
Fig.~\ref{fig:dia2} in the
limit of zero momenta for the external gluons. In this limit, the analytic properties
of the amplitudes are the same of those of the neutralino self-energy diagrams
with a quark--squark loop without external gluons attached (see
Fig.~\ref{fig:cutkoski}). For the self-energy diagrams, only one cut is
possible, namely when $\mc\ge m_{\tilde{Q}}+m_Q$, or $m_{\tilde{Q}} \le m_\chi - m_Q$. This corresponds to the opening
of the decay process $\chi\rightarrow \tilde{Q}+Q$. As a consequence, the amplitude is regular for $m_{\tilde{Q}}\ge \mc - m_Q$. This is
also explicitly derived in the Appendix.

The HIN decomposition of the
neutralino--gluon scattering loop  into long distance and
short distance contributions (see Eqs. (\ref{eq:hinfm})) allows to gain
more insight into the origin of the resonance at $m_{\tilde{Q}}=m_\chi-m_Q$. As
explained in HIN, the long--distance amplitudes
$f_{+}^{l}$ and $f_{-}^{l}$ originate from the diagram
of Fig.~\ref{fig:dia2}(a), while the
short--distance amplitudes $f_{+}^{s}$ and $f_{-}^{s}$ originate from the diagram
of Fig.~\ref{fig:dia2}(c)(notice that in the Fock-Schwinger gauge used by
HIN the other diagrams vanish). In both cases, when
$m_{\chi}\rightarrow m_b+m_{\tilde{b}}$, a color--singlet R--hadron can
be formed. In particular, denoting exotic color--triplet states by
$C_3$, color anti--triplet states by $C_{\bar{3}}$ and color octect
states by $C_8$ (as for instance in~\cite{r_hadrons}) the resonant
behavior in $f_{+}^{l}$, $f_{-}^{l}$ can be interpreted as the
formation of a $C_8qqq$ state, while the resonant behavior in
$f_{+}^{s}$, $f_{-}^{s}$ by the scattering of a $C_{3}\bar{q}$ or
$C_{\bar{3}}q$ state off a gluon in the proton (the analogous
scattering of $C_{3}\bar{q}$ or $C_{\bar{3}}q$ off a quark being
forbidden by color conservation). In Fig.~\ref{fig:long_and_short} the
components of these resonant states are grouped by boxes.

Figs.~\ref{fig:sigma_direct_a1_b0},~\ref{fig:sigma_direct_a0_b1},~\ref{fig:sigma_direct_a1_b1}
 show the neutralino--nucleon cross--section as a function of the sbottom mass for the representative choice
$m_{\chi}=10$ GeV, and for the cases
$(a_{\tilde{b}},b_{\tilde{b}})=(1,0)$, (0,1) and (1,1), respectively.  In each figure,
 the thick green solid line is the
neutralino--nucleon cross section calculated using
Eq.~(\ref{eq:cross_section}) with the transition amplitude $f$ given in
Eq.~(\ref{eq:fb}), the thin blue dotted line is the same quantity
with the transition amplitude from Ref.~\cite{drees_nojiri} (DN) in
Eq.~(\protect\ref{eq:dnfb2}) instead, while the red dashed line represents
the same cross section calculated with the transition amplitude given in
Eq.~(\protect\ref{eq:hinfb}) and taken from
Ref.~\cite{hisano_cross_section} (HIN). In the same figures, the gray
dot-dashed line represents the calculation in the heavy--squark limit
where the transition amplitude is given by the
Eq.~(\protect\ref{eq:heavy_squark}) modified by an ad--hoc substitution
$m_{\tilde{b}}^2\rightarrow m_{\tilde{b}}^2-(\mc+m_b)^2$ in the
propagator. (In Fig.\ref{fig:sigma_direct_a1_b1}, where
$a_{\tilde{b}}=b_{\tilde{b}}=1$, this last curve is missing because the
corresponding cross section vanishes).

It appears evident from Figs.~\ref{fig:sigma_direct_a1_b0},~\ref{fig:sigma_direct_a0_b1},~\ref{fig:sigma_direct_a1_b1}
that the prescription
$m_{\tilde{b}}^2\rightarrow m_{\tilde{b}}^2-(\mc+m_b)^2$ in the
propagator of the heavy--squark expression of
Eq. (\ref{eq:heavy_squark}) introduces a spurious pole at
$m_{\tilde{b}}=\mc+m_b$. The same spurious pole is contained in the second term
of the DN amplitude of Eq.~(\ref{eq:dnfb_quark_twist}) proportional to
the twist--two quark operator, but it is cured by the strategy adopted in
Ref.~\cite{drees_nojiri} of using in the cross section the smaller
amplitude between Eq.~(\ref{eq:dnfb_quark_twist})
and~(\ref{eq:dnfb_gluon_twist}), where in the latter expression the second
term is proportional to the twist--two gluon term instead. Notice,
however, that this approach may lead to somewhat erratic predictions
in the case Eq. (\ref{eq:dnfb_quark_twist}) is suppressed by some
accidental cancellation, as in Fig. \ref{fig:sigma_direct_a1_b0} where
this occurrence causes a spurious dip in the cross section. In all
other cases the predictions of Eqs. (\ref{eq:fb}) and (\ref{eq:dnfb2})
coincide. On the other hand, the spurious pole in the term
proportional to the twist--two quark operator is responsible for the
peak at $m_{\tilde{b}}=\mc$ in the HIN prediction of
Eq. (\ref{eq:hinfb}) (dashed line in
Figs. \ref{fig:sigma_direct_a1_b0}--\ref{fig:sigma_direct_a1_b1}), the
reason being that in that case the authors chose to take a vanishing
quark mass in the propagator (see Eq.(\ref{eq:his_gq})). Notice that
in HIN the $g^{(i,Q)\rm HIN}_G$ terms are
neglected on the ground that they are suppressed by $\alpha_s$ (see
Eqs.(\ref{eq:ourgG}) and
(\ref{eq:fpmgpm3}-\ref{eq:fpmgpm4})). However, it should be more
appropriate to say that HIN adopted the ``hybrid'' form of the
effective Lagrangian given in Eq. (\ref{eq:hybrid}) where the
off--trace contribution is expressed in terms of the twist--two quark
operator.  In this case including also the terms proportional to the
twist--two gluon operator would have implied a double counting.

A quantitative comparison between the neutralino--nucleon cross
section discussed so far and the output from the two popular
public codes \Mo~(discussed in Section \ref{sec:micromega}) and
DarkSUSY (discussed in Section \ref{sec:darksusy}) is provided in
Figs.~\ref{fig:sigma_a1_b0_ms3}--\ref{fig:sigma_a1_b1_ds}. In
particular, Figs. \ref{fig:sigma_a1_b0_ms3}, \ref{fig:sigma_a0_b1_ms3}
and \ref{fig:sigma_a1_b1_ms3} show the comparison with \Mo~ for the
cases
$(a_{\tilde{b}},b_{\tilde{b}})=(1,0)$, (0,1) and (1,1), respectively, while
Figs. \ref{fig:sigma_a1_b0_ms3}--\ref{fig:sigma_a1_b1_ms3} show the
same for DarkSUSY. In all these figures, for comparison we also show the green solid and
gray dot-dashed lines of
Figs.~\ref{fig:sigma_direct_a1_b0}--\ref{fig:sigma_direct_a1_b1} (for
the corresponding values of $a_b$ and $b_b$).

In Figs.~\ref{fig:sigma_a1_b0_ms3}--\ref{fig:sigma_a1_b1_ms3} the
orange thin solid line shows the output of \Mo\ for the option
``default'', while the orange dashed lines correspond to the same
quantity for the option ``FeScLoop.'' In both cases, the presence of
the spurious pole at $m_{\tilde{b}}=\mc+m_b$ is evident (for the
``default'' case, see Eqs. (\ref{eq:MO1}--\ref{eq:MO4}); for the
``FeScLoop'' option, it is not clear how a pole arises from
(\ref{eq:fescloop1}--\ref{eq:fescloop4}) ).

In Figs. \ref{fig:sigma_a1_b0_ds}--\ref{fig:sigma_a1_b1_ds} the thin
red solid line represents the DarkSUSY output with the option
``default'', the long red dashed line the DarkSUSY output with the
option ``pole'', and the short red dashed line the DarkSUSY output with
the option ``Drees--Nojiri''. With the exception of the option ''pole,'' which introduces a pole by hand, the unphysical pole at $m_{\tilde{b}}=\mc+m_b$
is not present.

From inspection of
Figs. \ref{fig:sigma_a1_b0_ms3}--\ref{fig:sigma_a1_b1_ds} we conclude
that all \mo\ 3.1 options have a spurious unphysical pole at $m_{\tilde{b}}=m_\chi+m_b$, while the default DarkSUSY option, although not showing a pole, fails to capture the full behavior of the cross section according to the DN and HIN one--loop calculation. Only by
selecting the ``Drees--Nojiri'' option in DarkSUSY the correct formula
is used, an option that however is not the current default in DarkSUSY 5.1 and may be missed by some users.

\section{Conclusions}
\label{sec:conclusions}
In this paper we have reviewed the neutralino--nucleon scattering
cross section when the neutralino mass $m_{\chi}$ is almost degenerate
with the sbottom mass $m_{\tilde{b}}$. We have shown that this
particular scenario, which has also been discussed in the literature
as a viable explanation of the experimental excesses observed by the
DAMA and CoGeNT experiments in terms of light WIMPs, may not be
properly accounted for by available calculation packages such as
DarkSUSY 5.1 in its default option and \mo\ 3.1.  In particular, we
have discussed the analytical continuation of the one--loop
gluon--neutralino scattering amplitude to the regime
$m_{\tilde{b}}<m_{\chi}$, showing that the neutralino--nucleon cross
section develops a pole when $m_{\chi}=m_{\tilde{b}}+m_b$.  This
feature is due to the fact that in the limit of vanishing gluon
momenta the loops describing the neutralino-gluon scattering have the
same analytic behavior as the neutralino self--energy with a quark and
a squark running in the loop, with a single cut when the decay process
$\chi\rightarrow \tilde{Q}+Q$ becomes kinematically accessible. Thus
when $\mc\le m_{\tilde{Q}}+m_Q$ the amplitude is analytic. The only
pole of the cross section can be further interpreted as the formation
of a resonant state in the nucleon, specifically, either an R-hadron
$C_8 qqq$~\cite{r_hadrons}, with $C_8$ a $b\tilde{b}$ color--octect
state and $q$ the valence quarks in the nucleon, or a color triplet
anti--triplet state $b\tilde{b}^*$ or $\bar{b}\tilde{b}$. These
resonant states are however kinematically not accessible if the
neutralino is the LSP and thus lighter than the sbottom, as is the
case for neutralino dark matter. Our analysis clearly shows that the
common practice of estimating the cross section by the substitution
$m_{\tilde{b}}^{-4}\rightarrow
[(m_{\chi}+m_b)^2-m_{\tilde{b}}^2]^{-2}$ in the propagator of an
effective four--fermion quark--neutralino interaction (also in the
coefficients multiplying twist--two quark terms) should be
discouraged, since it corresponds to adding a spurious pole to the
scattering cross section. The necessity to avoid such a pole was also
recognized in the work by DN (Ref.~\cite{drees_nojiri}). We also
pointed out that the very common practice of writing the effective
Lagrangian for the neutralino scattering through squark exchange as
the sum of a ``trace'' part constructed from the loop--induced
neutralino--gluon effective Lagrangian and an ``off-trace'' part
constructed from the twist--two neutralino--quark effective Lagrangian
is a hybrid approach with no robust justification. In the case of a
heavy quark (such as the $b$-quark) a description of the neutralino
scattering only in terms of a neutralino--gluon interaction appears
more consistent, in particular it avoids the need to describe the
cross section behavior in a semi--empirical way when $m_{\tilde{Q}}
\rightarrow \mc$.

\acknowledgments

P.G.\ was supported in part by the National Science Foundation
under Award PHY-1068111, and acknowledges the hospitality of CETUP 2013 where this work was completed. S.S. acknowledges support by the
National Research Foundation of Korea (NRF) grant funded by the Korea
government (MEST) (No.2012-0008534).

\appendix
\section{Appendix}

\def\ms{m_{\tilde{q}}}
\def\mq{m_{q}}
\def\mx{m_{\chi}}

Here we give the analytic expressions of the DN loop integrals contained
in Eq. (\ref{b_factors}) and of the HIN loop integrals in Eq.~(\ref{eq:hinfp})-(\ref{eq:hinfm}). We also show that these loop integrals are regular at $m_{\tilde{q}} = m_\chi + m_q$ and diverge at $m_{\tilde{q}} = m_\chi - m_q$.

Let
\begin{align}
D &= x^2 \mcsq + x \left( \msqsq- \mqsq - \mcsq \right) + \mqsq,
\label{b2a} \\
\Delta &= 2 \mcsq \left (\mqsq + \msqsq \right) - m^4_{\chi} -
\left( \msqsq - \mqsq \right)^2, \label{b2b} \\
L &= \frac {2} {\sqrt{|\Delta|}} \left[ \arctan \frac {\rt} {\mqsq + \msqsq - \mcsq}+\Theta(m_{\chi}^2-m_q^2-m_{\tilde{q}}^2)\pi \right],
\ \ \ \ \ \ \Delta \geq 0, \label{eq:l_delta_negative} \\
 &= \frac {1} {\sqrt{|\Delta|}} \left[ \ln \frac {\mqsq + \msqsq - \mcsq + \rt}
{\mqsq + \msqsq - \mcsq - \rt}+\Theta(m_{\chi}^2-m_q^2-m_{\tilde{q}}^2)2\pi i \right] \ \ \
\Delta \leq 0.
\label{b2c}
\end{align}
The terms proportional to the Heaviside step function $\Theta$ in
Eqs.~(\ref{eq:l_delta_negative})-(\ref{b2c}) extend the expression of $L$ given in DN and HIN to
$m_{\tilde{q}}<m_{\chi}$. In Eq.~(\ref{eq:l_delta_negative}), the $\arctan$ function is the principal arctan function with values in the range $(-\pi/2,\pi/2)$, and the $\Theta$ term is equivalent to taking a different branch of the $\arctan$ so that
it is continuous in the first and second quadrants, i.e.\ the range is $(0,\pi)$.

The DN loop integrals $I_n(\msq,m_q,\mc)$ are explicitly given by
\begin{align}
I_1(\msq,m_q,\mc) &= \int_0^1 dx \frac {x^2 - 2x + 2/3} {D^2} \label{b1a}\\
&= \frac {1} {\Delta} \left[ \frac {\mqsq-\mcsq}{3 \msqsq} - \frac {2}{3}
\frac {\msqsq - \mcsq}{\mqsq} - \frac{5}{3} + \left(2 \msqsq - \frac{2}{3}
\mcsq \right) L \right]; \nonumber
\end{align}
\begin{align}
I_2(\msq,m_q,\mc) &=  \int_0^1 dx \frac {x(x^2 - 2x + 2/3)} {D^2}
\nonumber \\
&= \frac {1}{2 m^4_{\chi}} \left[ \ln \frac {\msqsq} {\mqsq} - \left(
\msqsq -\mqsq - \mcsq \right) L \right] \nonumber \label{b1b} \\
&+ \frac {1} {\Delta} \left\{ \left[ \frac {m_q^4 - \mqsq \msqsq} {\mcsq}
- \frac {7}{3} \mqsq + \frac{2}{3} (\mcsq - \msqsq) \right] L \right.
\nonumber \\ & \left. \hspace*{1cm} + \frac {\mqsq -\mcsq} {3\msqsq}
+ \frac {\msqsq - \mqsq} {\mcsq} + \frac {2}{3} \right\}.
\end{align}
\begin{align}
I_3(\msq,m_q,\mc) &=  \int_0^1 dx \frac {x^2(1-x)^2} {D^3}
\nonumber\\
&=\frac{3(\mcsq-\mqsq-\msqsq)}{\Delta^2}+ \frac{L}{\Delta}
\left( -1 + \frac{6\mqsq\msqsq}{\Delta}\right) ;\label{b1c}
\end{align}
\begin{align}
I_4(\msq,m_q,\mc) &=  \int_0^1 dx \frac {x^3(1-x)^2} {D^3}
\nonumber \label{eq:i_4}\\
&=\frac{1}{2 m_{\chi}^6}\left[ \ln\frac{\msqsq}{\mqsq}-
(\msqsq-\mqsq-m^2_{\chi})L \right] - \frac{1}{\msqsq m_{\chi}^4}
\\
&-\frac{\mqsq(\msqsq-\mqsq-\mcsq)}{m_{\chi}^4 \Delta}L
+\frac{1}{\Delta}\left[ \frac{\mqsq}{m_{\chi}^4}
-\frac{1}{\msqsq}\left( 1-\frac{\mqsq}{\mcsq} \right)^2 + \frac{1}{2\mcsq}
\right]
\nonumber\\
&+\frac{3\mqsq}{\Delta^2}
\left\{ 1+\frac{\msqsq-\mqsq}{\mcsq}
+ \left[ \frac{\mqsq(\mqsq-\msqsq)}{\mcsq}-2\mqsq-\msqsq +\mcsq \right] L
\right\} ;  \nonumber
\end{align}
\begin{align}
I_5(\msq,m_q,\mc) &=  \int_0^1 dx \frac {x(1-x)(2-x)} {D^2}
\nonumber\\
&=\frac{1}{2 m_{\chi}^4}
\left[ \ln \frac{\msqsq}{\mqsq} -(\msqsq-\mcsq-\mqsq)L \right]
\\
&-\frac{1}{\Delta}\left\{ L \left[
2(\msqsq-\mcsq)+3\mqsq+\frac{\mqsq(\msqsq-\mqsq)}{\mcsq}
\right]
%\right. \nonumber\\ &\ \ \ \ \left.
-3+\frac{\mqsq-\msqsq}{\mcsq}\right\}. \nonumber
\end{align}
In the first line of the explicit expression of $I_4$, Eq.(\ref{eq:i_4}), we have corrected two typos that appear in Ref.~\cite{drees_nojiri}, namely the power of $m_\chi$ in the coefficient of $L$ in parenthesis, and the power of $m_\chi$ in the denominator of the second term.

HIN introduce the following loop integrals,
\begin{align}
B_0^{(n,m)} & = \int \frac{d^4q}{i\pi^2} \frac{1}{((p+q)^2-m_q^2)^n(q^2-m_{\tilde{q}}^2)^m} , \\
p_\mu B_1^{(n,m)} & = \int \frac{d^4q}{i\pi^2} \frac{q_\mu}{((p+q)^2-m_q^2)^n(q^2-m_{\tilde{q}}^2)^m} .
\end{align}
These integrals correspond to diagrams in which the four-momentum of the gluons is neglected and the four-momentum of the $\chi$ is $p$. An analytic calculation gives
\begin{align}
B_0^{(1,4)} & = -\frac{1}{3} \int_0^1 dx \frac{x^3}{D^3}, \\
B_1^{(1,4)} & = +\frac{1}{3} \int_0^1 dx \frac{(1-x)x^3}{D^3}, \\
B_0^{(4,1)} & = -\frac{1}{3} \int_0^1 dx \frac{(1-x)^3}{D^3}, \\
B_1^{(4,1)} & = +\frac{1}{3} \int_0^1 dx \frac{(1-x)^4}{D^3}, \\
B_0^{(3,1)} & = +\frac{1}{2} \int_0^1 dx \frac{(1-x)^2}{D^2}.
\end{align}

From either the DN or HIN expressions we find (notice that HIN's quantity $\Delta$ has the opposite sign to ours)
\begin{align}
f_{S}^{(Q)} & = \frac{\Delta (m_\chi^2-2m_{\tilde{q}}^2-m_q^2) - 6 m_{\tilde{q}}^2 m_q^2 (m_q^2-m_{\tilde{q}}^2-m_\chi^2) }{2 \Delta^2 m_{\tilde{q}}^2} + \frac{3m_{\tilde{q}}^2m_q^2 (m_q^2-m_{\tilde{q}}^2+m_\chi^2)}{\Delta^2} L ,
\\
f_{D}^{(Q)} & = \frac{3((m_{\tilde{q}}^2-m_q^2)^2-m_\chi^2(m_{\tilde{q}}^2+m_q^2) )}{\Delta^2} + \frac{m_q^4+m_q^2m_{\tilde{q}}^2-2m_{\tilde{q}}^4-m_q^2m_\chi^2+2m_{\tilde{q}}^2m_\chi^2}{2\Delta m_{\tilde{q}}^2 m_q^2}
\nonumber \\ & \quad
+ \frac{3m_{\tilde{q}}^2(2m_q^2m_\chi^2-\Delta)}{\Delta^2} L ,
\\
g_{S}^{(Q)} & =  \frac{\alpha_s}{4\pi m_\chi^4} \log\frac{m_{\tilde{q}}^2}{m_q^2}
\nonumber \\ & \quad + \frac{\alpha_s}{3\pi} \Bigg[ - \frac{3m_q^2(m_q^2-m_{\tilde{q}}^2-m_\chi^2)}{\Delta^2}  - \frac{2m_q^4-m_q^2m_{\tilde{q}}^2-m_{\tilde{q}}^4-4m_q^2m_\chi^2-4m_{\tilde{q}}^2m_\chi^2+2m_\chi^4}{2\Delta m_{\tilde{q}}^2m_\chi^2}
\nonumber \\ & \qquad\quad  -\frac{1}{m_{\tilde{q}}^2m_\chi^2}
+ L \Bigg( \frac{3(m_q^2-m_{\tilde{q}}^2+m_\chi^2)}{4m_\chi^4}
\nonumber \\ & \qquad\qquad\qquad\qquad\qquad
+ \frac{3m_q^4-3m_q^2m_{\tilde{q}}^2-m_q^2m_\chi^2-2m_{\tilde{q}}^2m_\chi^2+2m_\chi^4}{2\Delta m_\chi^2}
\nonumber \\ & \qquad\qquad\qquad\qquad\qquad
+ \frac{3m_q^2(m_q^4-m_q^2m_{\tilde{q}}^2-2m_q^2m_\chi^2-m_{\tilde{q}}^2m_\chi^2+m_\chi^4)}{\Delta^2}\Bigg) \Bigg] ,
\\
g_{D}^{(Q)} & = \frac{\alpha_s}{3\pi} \, \frac{m_\chi^2}{\Delta^2} \Big[ 3(m_\chi^2-m_q^2-m_{\tilde{q}}^2) + (6m_q^2m_{\tilde{q}}^2-\Delta) L  \Big]
\end{align}

One can easily see that there is no pole in $f_{S,D}^{(Q)}$ and $g_{S,D}^{(Q)}$ at $\msq=\mc+m_q$, directly from their expressions as integrals in $x$. In fact, at $\msq=\mc+m_q$, we have $D=(m_q+xm_\chi)^2$, which is never zero for $0\le x\le1$. Thus none of the integrals $I_k$ or $B_i^{(n,m)}$ has a singularity in the interval of integration. Hence they converge to a finite value.
Inserting $D=(m_q+xm_\chi)^2$ into the integrals gives
\begin{align}
\lim_{m_{\tilde{q}}\to m_\chi+m_q} f_{D}^{(Q)} = & \, - \frac{\mc(5m_q+3\mc)}{10m_q^3(\mc+m_q)^3} ,
\\
\lim_{m_{\tilde{q}}\to m_\chi+m_q}f_{S}^{(Q)} = & \, \frac{5m_q+\mc}{20m_q^2(\mc+m_q)^3} ,
\\
\lim_{m_{\tilde{q}}\to m_\chi+m_q}g_{D}^{(Q)} = & \, \frac{\alpha_s}{3\pi}\, \frac{m_\chi^2}{30m_q^3(m_\chi+m_q)^3},
\\
\lim_{m_{\tilde{q}}\to m_\chi+m_q}g_{S}^{(Q)} = & \, \frac{\alpha_s}{4\pi} \log\frac{(m_\chi+m_q)^2}{m_q^2}
\nonumber \\ & + \frac{\alpha_s}{3\pi}\, \frac{13m_\chi^4-15m_\chi^3 m_q-165 m_\chi^2m_q^2-225m_\chi m_q^3-90m_q^4}{60 m_q^2 m_\chi^3 (m_\chi+m_q)^3}.
\label{eq:no_pole}
\end{align}
There is no resonance at $\msq=\mc+m_q$.

Similarly, when $m_{\tilde{q}}=m_\chi$, we have $D=m_q^2(1-x)+m_\chi^2x^2$, which is positive for $m_\chi>2m_q$.

Instead, at $\msq=\mc-m_q$, we have $D=(m_q-x\mc)^2$, which vanishes inside the range of integration at $x=\xi\equiv m_q/\mc$ and is otherwise positive. Thus each integral diverges. An explicit calculation shows that there is no cancellation when combining the integrals. The dominant divergent parts as $m_{\tilde{q}} \to m_\chi - m_q$ are
\begin{align}
m_q f_{D}^{(q)} \simeq m_\chi f_{S}^{(q)} \simeq -\frac{3\pi}{\alpha_s} m_q g_{D}^{(q)} \simeq -\frac{3\pi}{\alpha_s} m_\chi g_{S}^{(q)} \simeq -\frac{\xi^3(1-\xi)^2}{m_\chi^3} \int_{-\xi}^{1-\xi} \frac{dy}{y^6} \to \infty.
\end{align}

\end{document}